%% file: main.tex
\pdfoutput=1
\documentclass[12pt,a4paper]{article}
\usepackage{ifthen} % for conditional statements
\newboolean{pdflatex}
\setboolean{pdflatex}{true} % False for eps figures 
\usepackage{float}

\newboolean{articletitles}
\setboolean{articletitles}{true} % False removes titles in references

\newboolean{uprightparticles}
\setboolean{uprightparticles}{false} %True for upright particle symbols

\def\paperauthors{
Christine A. Aidala$^{1}$,
Michael Buchar$^{2}$,
Chris Burr$^{2}$,
Dillon S. Fitzgerald$^{1}$,
Cameron Duncan McClymont$^{2}$,
Adam Morris$^{3}$,
Sebastian Neubert$^{4}$,
Piet Nogga$^{4}$,
Ryunosuke O'Neil$^{2}$,
Eduardo Rodrigues$^{5}$,
Daan Eduard Rosendal$^{6}$,
Tibor \v{S}imko$^{2}$\\
}

\def\paperaffiliations{\textit{\small{
$^{1}$University of Michigan, United States\\
$^{2}$CERN, Switzerland\\
$^{3}$Vilnius University, Lithuania\\
$^{4}$University of Bonn, Germany\\
$^{5}$Oliver Lodge Laboratory, University of Liverpool, Liverpool, United Kingdom\\
$^{6}$Now at Lakera company, Zurich, Switzerland\\
}}}
\date{May 2026}
\def\paperasciititle{The LHCb Ntupling Service} % Set ASCII title here !! MAKE sure it's only ASCII characters !! 
\def\paperkeywords{{High Energy Physics}, {LHCb}} % Comma separated list
\def\papercopyright{\the\year\ CERN for the benefit of the LHCb collaboration} % new since 9/Apr/2018
\def\paperlicence{CC BY 4.0 licence}
\def\paperlicenceurl{https://creativecommons.org/licenses/by/4.0/}

\newif\ifEnableSectionTOCLinks
\EnableSectionTOCLinksfalse % deactivated

\input{LHCb/preamble}
\usepackage{longtable} % only for template; not usually to be used in PAPERs
\usepackage{subcaption}
\usepackage{cleveref}
\usepackage{tikz}
\usetikzlibrary{shapes.geometric, positioning, arrows.meta}
\definecolor{boxblue}{RGB}{96, 140, 203}
\definecolor{boxorange}{RGB}{224, 130, 0}
\definecolor{boxgreen}{RGB}{46, 160, 67}
\definecolor{boxred}{RGB}{235, 0, 0}
\begin{document}

%%%%%%%%%%%%%%%%%%%%%%%%%
%%%%% Title     %%%%%%%%%
%%%%%%%%%%%%%%%%%%%%%%%%%
\renewcommand{\thefootnote}{\fnsymbol{footnote}}
\setcounter{footnote}{1}
\newcommand{\period}{.}

% %%%%%%% CHOOSE TITLE PAGE--------
%\onecolumn
\input{title-LHCb-PAPER}

%\twocolumn
% %%%%%%%%%%%%% ---------

\renewcommand{\thefootnote}{\arabic{footnote}}
\setcounter{footnote}{0}

%%%%%%%%%%%%%%%%%%%%%%%%%%%%%%%%
%%%%%  Table of Content   %%%%%%
%%%%%%%%%%%%%%%%%%%%%%%%%%%%%%%%
%%%% Uncomment if desired
%\tableofcontents

\cleardoublepage

%%%%%%%%%%%%%%%%%%%%%%%%%
%%%%% Main text %%%%%%%%%
%%%%%%%%%%%%%%%%%%%%%%%%%

\pagestyle{plain} % restore page numbers for the main text
\setcounter{page}{1}
\pagenumbering{arabic}

%% Uncomment during review phase. 
%% Comment before a final submission.
%\linenumbers

%% This is the main body
%% It is useful to have a single file so comments are not missed in overleaf.
\input{body}

\input{acknowledgements}

\addcontentsline{toc}{section}{References}
%\setboolean{inbibliography}{true}
\bibliographystyle{LHCb/LHCb}
\bibliography{main,LHCb/standard,LHCb/LHCb-PAPER,LHCb/LHCb-CONF,LHCb/LHCb-DP,LHCb/LHCb-TDR,LHCb/LHCb-PUB}

\newpage

\end{document}

%% file: LHCb/preamble.tex
\usepackage[top=1in, bottom=1.25in, left=1in, right=1in]{geometry}
\usepackage{microtype}
\usepackage{lineno}  % for line numbering during review
\usepackage{xspace} % To avoid problems with missing or double spaces after
\usepackage{caption} %these three command get the figure and table captions automatically small

\usepackage{graphicx}  % to include figures (can also use other packages)
\usepackage{color}
\usepackage{colortbl}
\graphicspath{{./figs/}} % Make Latex search fig subdir for figures
\usepackage{amsmath} % Adds a large collection of math symbols
\usepackage{amssymb}
\usepackage{amsfonts}
\usepackage{upgreek} % Adds in support for greek letters in roman typeset

\usepackage[normalem]{ulem} % for \sout strikeout in template tables

\newcommand*\patchAmsMathEnvironmentForLineno[1]{%
\expandafter\let\csname old#1\expandafter\endcsname\csname #1\endcsname
\expandafter\let\csname oldend#1\expandafter\endcsname\csname
end#1\endcsname
 \renewenvironment{#1}%
   {\linenomath\csname old#1\endcsname}%
   {\csname oldend#1\endcsname\endlinenomath}%
}
\newcommand*\patchBothAmsMathEnvironmentsForLineno[1]{%
  \patchAmsMathEnvironmentForLineno{#1}%
  \patchAmsMathEnvironmentForLineno{#1*}%
}
\AtBeginDocument{%
\patchBothAmsMathEnvironmentsForLineno{equation}%
\patchBothAmsMathEnvironmentsForLineno{align}%
\patchBothAmsMathEnvironmentsForLineno{flalign}%
\patchBothAmsMathEnvironmentsForLineno{alignat}%
\patchBothAmsMathEnvironmentsForLineno{gather}%
\patchBothAmsMathEnvironmentsForLineno{multline}%
\patchBothAmsMathEnvironmentsForLineno{eqnarray}%
}

\usepackage[pdftex,
            pdfauthor={\paperauthors},
            pdftitle={\paperasciititle},
            pdfkeywords={\paperkeywords}]{hyperref}
\usepackage{hyperxmp}
\hypersetup{
    pdfcopyright={Copyright (C) \papercopyright},
    pdflicenseurl={\paperlicenceurl}
}
\usepackage[colorinlistoftodos,textsize=scriptsize]{todonotes}

\usepackage[bottom,flushmargin,hang,multiple]{footmisc}

\usepackage[all]{hypcap} % Internal hyperlinks to floats.

\input{LHCb/lhcb-symbols-def} % Add in the predefined LHCb symbols

\hypersetup{
  colorlinks   = true, %Colours links instead of ugly boxes
  urlcolor     = blue, %Colour for external hyperlinks
  linkcolor    = blue, %Colour of internal links
  citecolor    = red   %Colour of citations
}

\ifEnableSectionTOCLinks
    \usepackage[explicit]{titlesec} % to change headings
    
    \let\oldcontentsline\contentsline
    \renewcommand

    \titleformat{\section}{\normalfont\Large\bf}{\hyperlink{tocsection.\thesection}{{\thesection} \parbox[t]{\dimexpr\textwidth-1pc}{#1}}}{1pc}{}

    \titleformat{\subsection}{\normalfont\bf}{\hyperlink{tocsubsection.\thesubsection}{{\thesubsection} \parbox[t]{\dimexpr\textwidth-1pc}{#1}}}{1pc}{}

    \titleformat{name=\section,numberless}[display]{}{}{0pt}{\normalfont\Huge\bfseries #1}
\fi

\usepackage{cite} % Allows for ranges in citations
\usepackage{LHCb/mciteplus}

%% file: LHCb/lhcb-symbols-def.tex
\usepackage{xspace} 
\usepackage{upgreek}

\def\lhcb   {\mbox{LHCb}\xspace}

\def\MagUp {\mbox{\em Mag\kern -0.05em Up}\xspace}

\ifthenelse{\boolean{uprightparticles}}%
{

 \def\Ppsi        {\ensuremath{\uppsi}\xspace}

 \def\PDelta      {\ensuremath{\Delta}\xspace}                 
 \def\PXi         {\ensuremath{\Xi}\xspace}                 
 \def\PLambda     {\ensuremath{\Lambda}\xspace}                 
 \def\PSigma      {\ensuremath{\Sigma}\xspace}                 
 \def\POmega      {\ensuremath{\Omega}\xspace}                 
 \def\PUpsilon    {\ensuremath{\Upsilon}\xspace}
 \let\oldPi\Pi
 \def\PPi         {\ensuremath{\oldPi}\xspace}

 \def\PB      {\ensuremath{\mathrm{B}}\xspace}                 
 \def\PD      {\ensuremath{\mathrm{D}}\xspace}                 
 \def\PJ      {\ensuremath{\mathrm{J}}\xspace}                 
 \def\PK      {\ensuremath{\mathrm{K}}\xspace}                 
 \def\Ps      {\ensuremath{\mathrm{s}}\xspace}

 \def\thebaroffset{0.0em}
}
{

 \def\Ppsi        {\ensuremath{\psi}\xspace}                 
                  
 \mathchardef\PDelta="7101
 \mathchardef\PXi="7104
 \mathchardef\PLambda="7103
 \mathchardef\PSigma="7106
 \mathchardef\POmega="710A
 \mathchardef\PUpsilon="7107
 \mathchardef\PPi="7105
 \def\PB      {\ensuremath{B}\xspace}                 
 \def\PD      {\ensuremath{D}\xspace}                 
 \def\PJ      {\ensuremath{J}\xspace}                 
 \def\PK      {\ensuremath{K}\xspace}                 
 \def\Ps      {\ensuremath{s}\xspace}

 \def\thebaroffset{0.18em}
}
\newcommand{\offsetoverline}[2][\thebaroffset]{\kern #1\overline{\kern -#1 #2}}%

\makeatletter
\ifcase \@ptsize \relax% 10pt
  \newcommand{\miniscule}{\@setfontsize\miniscule{4}{5}}% \tiny: 5/6
\or% 11pt
  \newcommand{\miniscule}{\@setfontsize\miniscule{5}{6}}% \tiny: 6/7
\or% 12pt
  \newcommand{\miniscule}{\@setfontsize\miniscule{5}{6}}% \tiny: 6/7
\fi
\makeatother

\DeclareRobustCommand{\optbar}[1]{\shortstack{{\miniscule (\rule[.5ex]{1.25em}{.18mm})}
  \\ [-.7ex] $#1$}}

\def\squark    {{\ensuremath{\Ps}}\xspace}

\def\kaon    {{\ensuremath{\PK}}\xspace}
\def\KorKbar {\kern \thebaroffset\optbar{\kern -\thebaroffset \PK}{}\xspace}

\def\Kpm     {{\ensuremath{\kaon^\pm}}\xspace}

\def\D       {{\ensuremath{\PD}}\xspace}

\def\DorDbar {\kern \thebaroffset\optbar{\kern -\thebaroffset \PD}\xspace}

\def\Dp      {{\ensuremath{\D^+}}\xspace}
\def\Dm      {{\ensuremath{\D^-}}\xspace}

\def\DpDm    {\ensuremath{\Dp {\kern -0.16em \Dm}}\xspace}

\def\B       {{\ensuremath{\PB}}\xspace}

\def\BorBbar {\kern \thebaroffset\optbar{\kern -\thebaroffset \PB}\xspace}

\def\Bd      {{\ensuremath{\B^0}}\xspace}

\def\BdorBdbar {\kern \thebaroffset\optbar{\kern -\thebaroffset \Bd}\xspace}

\def\Bpm     {{\ensuremath{\B^\pm}}\xspace}

\def\Bs      {{\ensuremath{\B^0_\squark}}\xspace}

\def\BsorBsbar {\kern \thebaroffset\optbar{\kern -\thebaroffset \Bs}\xspace}

\def\jpsi     {{\ensuremath{{\PJ\mskip -3mu/\mskip -2mu\Ppsi}}}\xspace}

\def\Y#1S{\ensuremath{\PUpsilon{(#1S)}}\xspace}

\def\LorLbar     {\kern \thebaroffset\optbar{\kern -\thebaroffset \PLambda}\xspace}

\def\to                 {\ensuremath{\rightarrow}\xspace}

\def\AT#1     {\ensuremath{A_{\mathrm{T}}^{#1}}\xspace}           % 2

\def\C#1      {\ensuremath{\mathcal{C}_{#1}}\xspace}                       % 9
\def\Cp#1     {\ensuremath{\mathcal{C}_{#1}^{'}}\xspace}                    % 7
\def\Ceff#1   {\ensuremath{\mathcal{C}_{#1}^{\mathrm{(eff)}}}\xspace}        % 9  
\def\Cpeff#1  {\ensuremath{\mathcal{C}_{#1}^{'\mathrm{(eff)}}}\xspace}       % 7
\def\Ope#1    {\ensuremath{\mathcal{O}_{#1}}\xspace}                       % 2
\def\Opep#1   {\ensuremath{\mathcal{O}_{#1}^{'}}\xspace}                    % 7

\newcommand{\aunit}[1]{\ensuremath{\text{\,#1}}}       
\newcommand{\tev}{\aunit{Te\kern -0.1em V}\xspace}
\newcommand{\gev}{\aunit{Ge\kern -0.1em V}\xspace}
\newcommand{\mev}{\aunit{Me\kern -0.1em V}\xspace}
\newcommand{\kev}{\aunit{ke\kern -0.1em V}\xspace}
\newcommand{\ev}{\aunit{e\kern -0.1em V}\xspace}
 
\newcommand{\mevc}{\ensuremath{\aunit{Me\kern -0.1em V\!/}c}\xspace}
\newcommand{\gevc}{\ensuremath{\aunit{Ge\kern -0.1em V\!/}c}\xspace}
\newcommand{\mevcc}{\ensuremath{\aunit{Me\kern -0.1em V\!/}c^2}\xspace}
\newcommand{\gevcc}{\ensuremath{\aunit{Ge\kern -0.1em V\!/}c^2}\xspace}
\def\gsim{{~\raise.15em\hbox{$>$}\kern-.85em
          \lower.35em\hbox{$\sim$}~}\xspace}
\def\lsim{{~\raise.15em\hbox{$<$}\kern-.85em
          \lower.35em\hbox{$\sim$}~}\xspace}

\def\tell1  {TELL1\xspace}
\def\ukl1   {UKL1\xspace}

\newcommand{\lhcborcid}[1]{\href{https://orcid.org/#1}{\hspace*{0.1em}\raisebox{-0.45ex}{\includegraphics[width=1em]{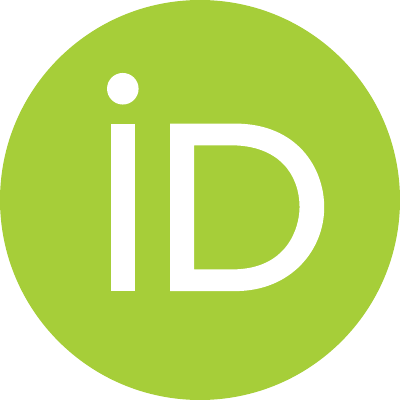}}}}

%% file: title-LHCb-PAPER.tex
% ===============================================================================
% Purpose: LHCb-PAPER journal paper title page template
% Author: 
% Created on: 2010-09-25
% ===============================================================================

%%%%%%%%%%%%%%%%%%%%%%%%%
%%%%%  TITLE PAGE  %%%%%%
%%%%%%%%%%%%%%%%%%%%%%%%%
\begin{titlepage}
\pagenumbering{roman}

% Header ---------------------------------------------------
\vspace*{-1.5cm}
\centerline{\large EUROPEAN ORGANIZATION FOR NUCLEAR RESEARCH (CERN)}
\vspace*{1.5cm}
\noindent
\begin{tabular*}{\linewidth}{lc@{\extracolsep{\fill}}r@{\extracolsep{0pt}}}
\ifthenelse{\boolean{pdflatex}}% Logo format choice
{\vspace*{-1.5cm}\mbox{\!\!\!\includegraphics[width=.14\textwidth]{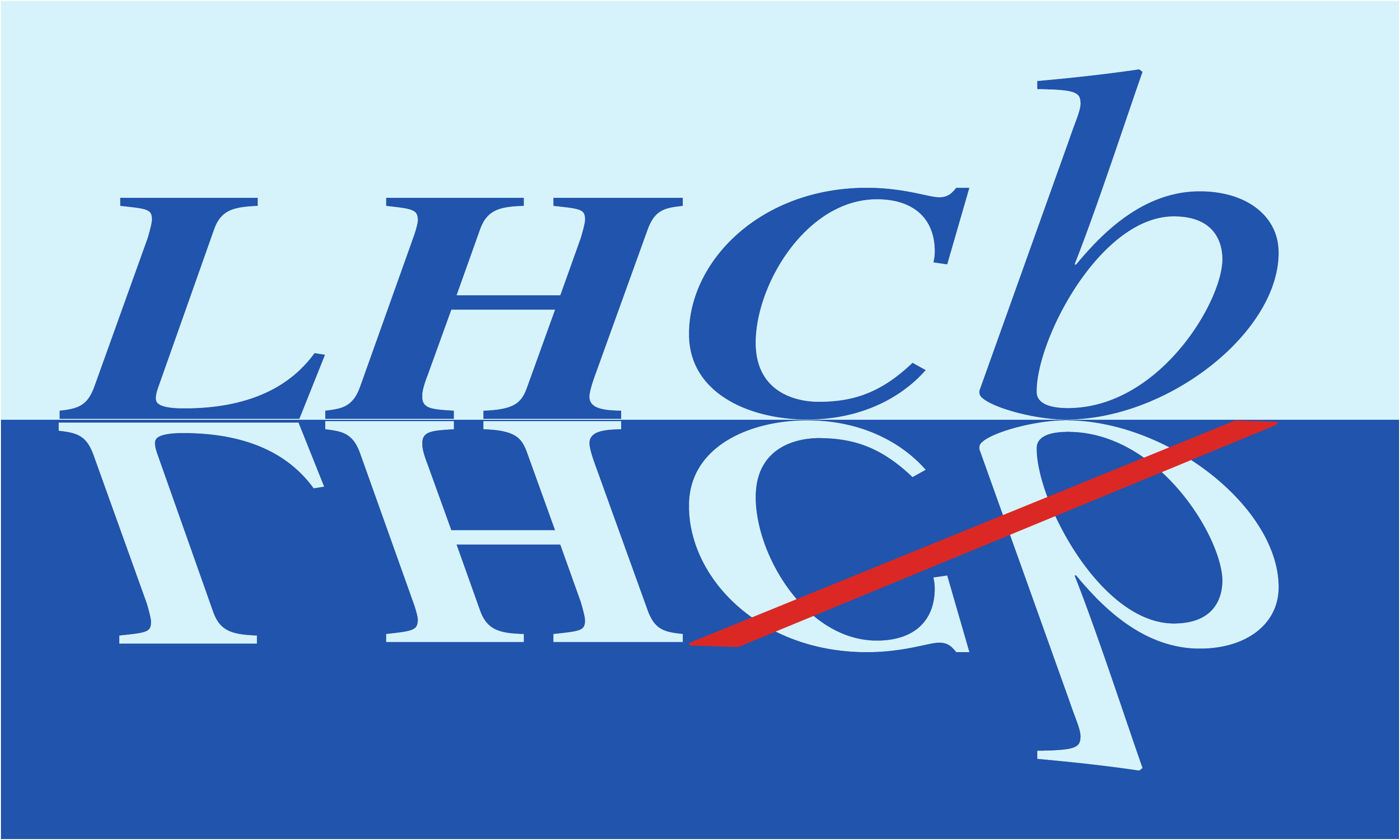}} & &}%
{\vspace*{-1.2cm}\mbox{\!\!\!\includegraphics[width=.12\textwidth]{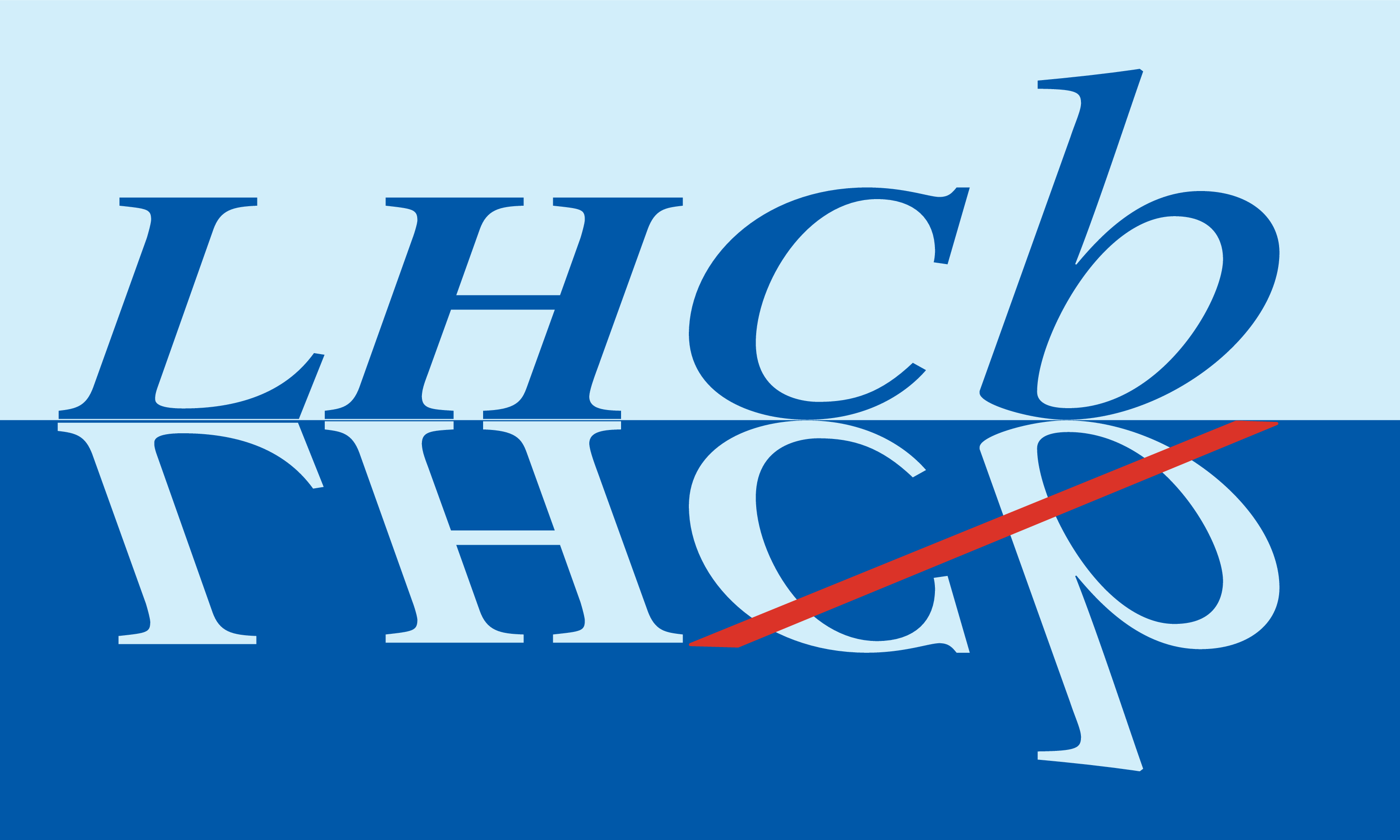}} & &}%
\\
% & & CERN-EP-2026-ZZZ \\  % ID 
% & & LHCb-PAPER-2026-YYY \\  % ID 
 & & \today \\ % Date - Can also hardwire e.g.: 23 March 2010
 & & \\
% not in paper \hline
\end{tabular*}

\vspace*{4.0cm}

% Title --------------------------------------------------
{\normalfont\bfseries\boldmath\huge
\begin{center}
% DO NOT EDIT HERE. Instead edit macro in main.tex to keep metadata correct
Open Science with the\\LHCb Ntupling Service
\end{center}
}

\vspace*{2.0cm}

% Authors -------------------------------------------------
\begin{center}

% Edit macro in main.tex to keep metadata correct

\paperauthors
\vspace*{0.4cm}
\paperaffiliations

\end{center}

\vspace{\fill}
% Abstract -----------------------------------------------
\begin{abstract}
\vspace{0.5cm}
\noindent
The data samples collected at the Large Hadron Collider at CERN are highly valuable, and the analysis of these data has led to numerous discoveries and the 2025 Breakthrough Prize in Fundamental Physics. Preserving and providing open access to these data for future analysts is imperative to maximize the scientific output of the collected data. The LHCb collaboration has developed a novel approach to this pursuit of open science, the LHCb Ntupling Service: a web service for on-demand production and publishing of custom LHCb open data, providing open access to both Run 1, and for the first time, Run 2 proton-proton collision data collected by the LHCb experiment, amounting to over 4 PB of data to explore.  
  
\end{abstract}

\vspace*{1.0cm}

\begin{center}
  Submitted to
  EPJ Research Infrastructures \
\end{center}

\vspace{\fill}

{\footnotesize 
% Edit macro in main.tex to keep metadata correct
\centerline{\copyright~\papercopyright. \href{\paperlicenceurl}{\paperlicence}.}}
\vspace*{2mm}

\end{titlepage}

%%%%%%%%%%%%%%%%%%%%%%%%%%%%%%%%
%%%%%  EOD OF TITLE PAGE  %%%%%%
%%%%%%%%%%%%%%%%%%%%%%%%%%%%%%%%

%  empty page follows the title page ----
\newpage
\setcounter{page}{2}
\mbox{~}
%\newpage
%
%% Author List ----------------------------
%%  You need to get a new author list!
%\input{LHCb_authorlist.tex}
%
%The author list for journal publications is provided by the Membership Committee shortly after 'approval to go to paper' has been given.
%%It will be made available on the page
%%\verb!http://www.physik.uzh.ch/~strauman/forMemCo/LHCb-PAPER-XXXX-XXX/! .
%It will be sent to you by email shortly after a paper number has beens assigned.
%The author list should be included already at first circulation, 
%to allow new members of the collaboration to verify whether they have been included correctly.
%Occasionally a misspelled name is corrected or associated institutions become full members.
%In that case, a new author list will be sent to you.
%In case line numbering doesn't work well after including the authorlist, try moving the \verb!\bigskip! after the last author to a separate line.
%
%
%The authorship for Conference Reports should be ``The LHCb
%  collaboration'', with a footnote giving the name(s) of the contact
%  author(s), but without the full list of collaboration names.

%% file: body.tex
\section{Introduction}
    Contemporary research in high energy physics (HEP) necessitates vast data volumes stored in custom file formats to meet the needs of experimental programs. This typically involves experiment-specific software to parse the custom file formats and extract necessary information for physics analysis. Both the complexity and sheer volumes of these datasets make it challenging to preserve and provide open access to these data. For example, all four main Large Hadron Collider (LHC) experiments have differing computing models and resulting file formats~\cite{LHCb-TDR-011, LHCB-TDR-017, CMS_computing_TDR, Mini_AOD, ATLAS_computing_TDR, ATLAS_CSW_Run3, ALICE_computing_TDR, ALICE_AO2D_TDR}, underlining the intricacies of the data preservation efforts in HEP. Nevertheless, data preservation efforts are crucial to maximize the physics potential of the collected samples. For this, CERN is committed to providing open access to these data as outlined in the CERN Open Data Policy~\cite{CERN_Open_Data_Policy}, encouraging adoption of a consistent approach between the four LHC experiments to publish scientific data in line with standards for Findability, Accessibility, Interoperability, and Reusability (FAIR)~\cite{FAIR}. 

    At the end of 2022, \lhcb announced the \lhcb Ntuple Wizard concept~\cite{lhcb_ntuple_wizard_paper}, a web application for configuring ntupling jobs for the \lhcb Analysis Productions batch processing system~\cite{lhcb_analysis_productions_paper} to be run on the Worldwide LHC Computing Grid (WLCG). In this case, ntupling refers to processing and storing experimental data into a columnar data format suitable for analysis with standard tools without the need for experiment-specific software. Significant progress has led to the development of an overarching system, the \lhcb Ntupling Service~\cite{lhcb_ntupling_service, service_proceedings}, to seamlessly handle the interface between the LHCb Ntuple Wizard and Analysis Productions, with dedicated databases for details of user requests and resulting ntuples from production jobs. It is an all-in-one web service for users to request custom ntuple productions, track the status of the ntupling jobs, communicate with the open data team, and retrieve the resulting ntuples. Internally, ntuple production at \lhcb is handled by the \texttt{DaVinci} application~\cite{LHCb-TDR-011,LHCB-TDR-017,DaVinci_git} through Python scripts with specific syntax related to the \lhcb data format. A noteworthy advantage of the \lhcb Ntupling Service is that configurations needed to steer the \texttt{DaVinci} application are captured through the Ntuple Wizard web interface, removing the necessity of learning \texttt{DaVinci} to produce ntuples. Throughout the paper, the LHCb Ntupling Service will often be referred to as the ``service'', while the \lhcb Ntuple Wizard will often be referred to as the ``wizard''.
    
\subsection {Current Open Data Landscape}
    The four large LHC experiments, in addition to many other HEP experiments, have efforts to provide open access to their collected data, each with their own computing model, leading to a complex landscape of data formats to support. Open data from HEP experiments are typically provided to users in one of two ways:
    \begin{enumerate}
        \item Experiment-specific data formats, along with the necessary software and instructions to analyze these data.
        \item A derived data format, such as a ROOT~\cite{Root,Antcheva:2009zz} ntuple\footnote{For each physics object or event of interest, a tuple of information is saved with n columns corresponding to n different physical quantities.}, removing the need for experiment-specific software.
    \end{enumerate}
    
    Both approaches have their advantages. Approach (1) offers additional flexibility and customization at the cost of a significantly steeper learning curve, requiring custom software and tools, as well as sufficient CPU to parse the data. Approach (2) is typically limited to particular selections meant to isolate physics processes of interest, ultimately up to the collaboration releasing the data. While this is excellent in terms of reducing the barrier of entry to work with these data, the fine-grained choices of physics objects and selections provided by collaborations can stifle creativity in research using open data. On the other hand, the approach in (1) comes at a significantly higher resource cost, as full data sets stored in experiment-specific file formats typically have large data volumes that need to be duplicated for public access. In December 2023, \lhcb released the entire Run 1 dataset to the CERN Open Data Portal (ODP)~\cite{CERN_ODP}, amounting to about 800\,TB of collision data in custom file formats, corresponding to approach (1). Applying the same methodology to the Run 2 data would amount to roughly 10 PB of custom file formats copied to the CERN Open Data Portal, which was not feasible from the standpoint of resource availability.
    
    It is worth mentioning the significant efforts by other HEP experiments to generalize the approach in (2) to provide a derived data format rich in high-level event information suitable for a number of physics analyses, without the need for experiment-specific software. Both the CMS and ATLAS collaborations at the LHC have efforts in this direction with their so-called NanoAOD~\cite{NANOAOD} and PHYSLITE~\cite{PHYSLITE} data formats, respectively, both with accompanying open data releases~\cite{ATLAS_OD_release, using-CMS-OD}. However, reduction of average event size is typically accomplished by isolating particular physics objects and/or skimming away event-level information. 

    The novel web service presented in this paper introduces an innovative methodology for providing open access to data collected in HEP experiments. The \lhcb Ntupling Service allows for on-demand production of custom \lhcb open data, offering exploration of the \lhcb data streams, specification of physics object selection, and customization of information to be written in the resulting ntuple(s). This offers both flexibility in terms of exploring physics selections, and a lower learning curve without the need for experiment-specific software to process the resulting ntuples, capturing the best of approaches (1) and (2). It also results in significantly smaller data volumes on the CERN Open Data Portal, as produced ntuples are typically on the order of a few to hundreds of GB.

     The paper is organized as follows: the architecture of the \lhcb Ntupling Service is discussed in~\Cref{sec:architecture}, followed by a detailed look at the user experience in~\Cref{sec:user-experience}.
     The integration with the CERN Open Data Portal is presented in~\Cref{sec:ODP-integration}, followed by a discussion of future plans / outlook in~\Cref{sec:future-outlook}, and finally, a summary in Section~\ref{sec:summary}.

\section{Architecture}
    \label{sec:architecture}

\begin{figure}[htbp]
        \centering
        \includegraphics[width=0.95\linewidth]{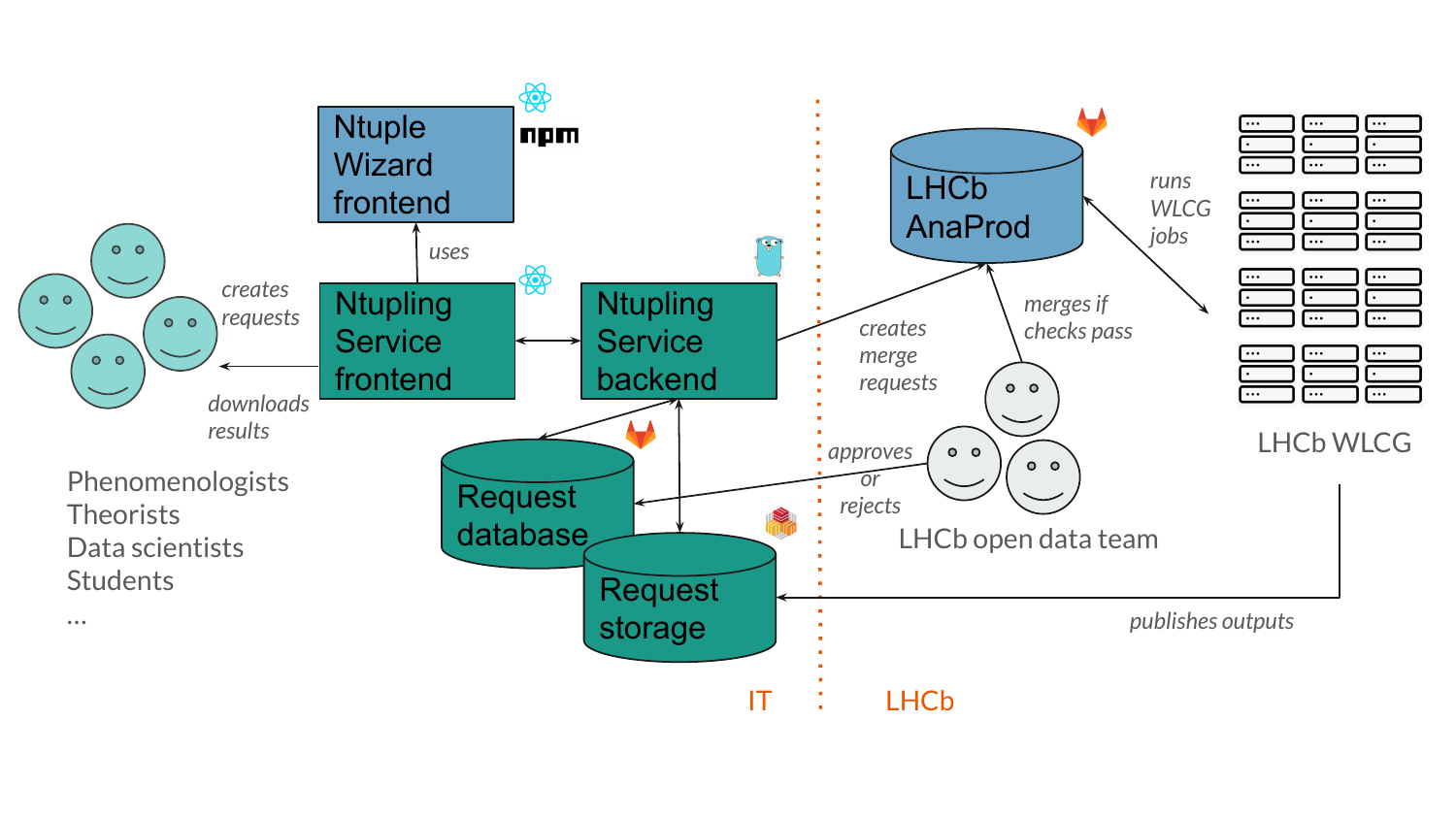}
        \caption{Architecture of the LHCb Ntupling Service, integrating the LHCb Ntuple Wizard and the Analysis Productions batch processing system. The rectangles represent code repositories while the cylinders represent databases, with light blue and green shading representing components managed by LHCb and CERN IT, respectively. The logos identify the technologies used for the different components, and the orange dashed line represents the operational division between LHCb and CERN IT. While LHCb is responsible for maintaining the wizard, CERN IT handles the integration with the service.}
        \label{fig:service_architecture}
    \end{figure}

The \lhcb Ntupling Service comprises three code repositories: a React~\cite{reactjs} \textbf{frontend} user interface, a Representational State Transfer (REST) \textbf{backend}, and a \textbf{DevOps} repository holding the deployment tooling and Helm chart, all privately hosted on the open GitLab code repository platform~\cite{gitlab}. A fourth component, the \textbf{\lhcb Ntuple Wizard}~\cite{lhcb_ntuple_wizard_paper}, is consumed by the frontend and is core to the building and submission of ntupling requests (see Section~\ref{sec:arch_wizard}). The \lhcb Analysis Productions batch processing system~\cite{lhcb_analysis_productions_paper} is the engine behind running production jobs on the WLCG to fulfill the requests, which are themselves stored as GitLab issues in a private repository comprising the requests database.  A diagram of the system architecture is shown in Figure~\ref{fig:service_architecture}.

    \subsection{Component Topology}

    The service has a three-layer structure:
    \begin{itemize}
        \item Incoming traffic first reaches a CERN Single Sign-On (SSO) proxy, which handles authentication and forwards requests to the frontend. Non-CERN users have options to sign in with a guest email, their organization (eduGain), or with a social account such as Google, GitHub, Facebook, Apple, or LinkedIn. CERN users can simply sign in using their CERN account.
        \item The frontend is a React web application that uses Nginx~\cite{nginx} to serve files and reverse-proxy \texttt{/api} requests to the backend.
        \item The backend is a Go~\cite{golang2022} server that uses GitLab to store and track requests.
    \end{itemize}

    Each ntupling request is a GitLab issue in a private repository. The issue's labels encode the request status, the description includes details about the requester, and the comment thread facilitates discussion between the requester and the \lhcb team. This design outsources the state management, audit trail, and human review workflow to GitLab's existing infrastructure rather than writing our own bespoke system.

    \subsection{Ntuple Wizard Integration}
    \label{sec:arch_wizard}

    The request creation flow is provided by the \textbf{\lhcb Ntuple Wizard}, a separate component embedded into the frontend via its npm package (\texttt{lhcb-ntuple-wizard})~\cite{lhcb_ntuple_wizard_npm, npm}. The wizard itself is hosted on GitLab~\cite{lhcb_ntuple_wizard_git}, where it is tagged and versioned for corresponding releases on npm. This keeps the logic of assembling a valid ntuple configuration decoupled from the request management logic, and allows for independent development and reuse of the wizard.

    The wizard guides the user through selecting datasets and physics objects of interest (e.g. decay candidates), and generates the necessary \texttt{YAML} files to steer the production configuration client-side. Upon submission, it sends the generated files to the backend, where they seed the request issue. After review from the \lhcb Open Data team, the committed configuration is transferred to Analysis Productions to fulfill the request.

    \subsection{Storage}

        Two storage systems support the service:
    
        \begin{itemize}
            \item A \textbf{CephFS} file share~\cite{CephFS} for user profiles, comments, and bookkeeping; and
            \item A set of \textbf{EOS directories} for the large ntuple output files and the gallery of curated example requests,
            where EOS is CERN's multi-PB disk-only storage technology~\cite{EOS}.
        \end{itemize}

    \subsection{Request Lifecycle}

\begin{figure}[htbp]
        % Generated by CERN's gemini instance
            \centering
            \resizebox{0.95\linewidth}{!}{%
            \begin{tikzpicture}[
                node distance=0.8cm and 0.8cm,
                font=\sffamily\small,
                basebox/.style={
                    draw=black,
                    rounded corners=5pt,
                    text=white,
                    align=center,
                    minimum width=3.6cm,
                    minimum height=1.0cm,
                    inner sep=4pt
                },
                bluebox/.style={basebox, fill=boxblue},
                orangebox/.style={basebox, fill=boxorange},
                greenbox/.style={basebox, fill=boxgreen},
                redbox/.style={basebox, fill=boxred, minimum width=2.4cm},
                decision/.style={
                    diamond,
                    draw=black,
                    aspect=2,
                    align=center,
                    inner sep=2pt,
                    fill=white,
                    minimum width=2.6cm
                },
                arrow/.style={-{Stealth[length=2.5mm, width=2mm]}, semithick}
            ]
        
                % --- Left Column ---
                \node [bluebox] (created) {Created};
                \node [orangebox, below=of created] (lhcb) {Awaiting LHCb review};
                \node [decision, below=of lhcb] (dec1) {Approved?};
                \node [redbox, right=of dec1] (rejected) {Rejected};
                \node [greenbox, below=of dec1] (app_test) {Approved for test\\production};
                \node [bluebox, below=of app_test] (run_test) {Running test\\production};
                \node [decision, below=of run_test] (dec2) {Succeeded?};
                \node [redbox, right=of dec2] (failed1) {Failed};
        
                % --- Right Column (Positioned for full clearance) ---
                \node [orangebox, right=5.2cm of created] (confirm) {Awaiting requester\\confirmation};
                \node [decision, below=of confirm] (dec3) {Confirmed?};
                \node [redbox, right=of dec3] (cancelled) {Cancelled};
                \node [greenbox, below=of dec3] (conf_real) {Confirmed for real\\production};
                \node [bluebox, below=of conf_real] (run_real) {Running real\\production};
                \node [decision, below=of run_real] (dec4) {Succeeded?};
                \node [redbox, right=of dec4] (failed2) {Failed};
                \node [greenbox, below=of dec4] (completed) {Completed};
        
                % --- Left Column Connections ---
                \draw [arrow] (created) -- (lhcb);
                \draw [arrow] (lhcb) -- (dec1);
                \draw [arrow] (dec1) -- node[above, pos=0.35, font=\sffamily\footnotesize] {No} (rejected);
                \draw [arrow] (dec1) -- node[right, pos=0.35, font=\sffamily\footnotesize] {Yes} (app_test);
                \draw [arrow] (app_test) -- (run_test);
                \draw [arrow] (run_test) -- (dec2);
                \draw [arrow] (dec2) -- node[above, pos=0.35, font=\sffamily\footnotesize] {No} (failed1);
        
                % --- Transition Line (Middle Channel Path) ---
                \draw [arrow] (dec2.south) 
                    -- node[right, pos=0.15, font=\sffamily\footnotesize] {Yes} ++(0,-0.8cm)
                    -| ([xshift=-0.8cm]confirm.west)
                    |- ([yshift=0.8cm]confirm.north)
                    -- (confirm.north);
        
                % --- Right Column Connections ---
                \draw [arrow] (confirm) -- (dec3);
                \draw [arrow] (dec3) -- node[above, pos=0.35, font=\sffamily\footnotesize] {No} (cancelled);
                \draw [arrow] (dec3) -- node[right, pos=0.35, font=\sffamily\footnotesize] {Yes} (conf_real);
                \draw [arrow] (conf_real) -- (run_real);
                \draw [arrow] (run_real) -- (dec4);
                \draw [arrow] (dec4) -- node[above, pos=0.35, font=\sffamily\footnotesize] {No} (failed2);
                \draw [arrow] (dec4) -- node[right, pos=0.35, font=\sffamily\footnotesize] {Yes} (completed);
        
            \end{tikzpicture}%
            }
            \caption{Flowchart of GitLab issue labels used to process user requests.}
            \label{fig:gitlab_labels}
        \end{figure}
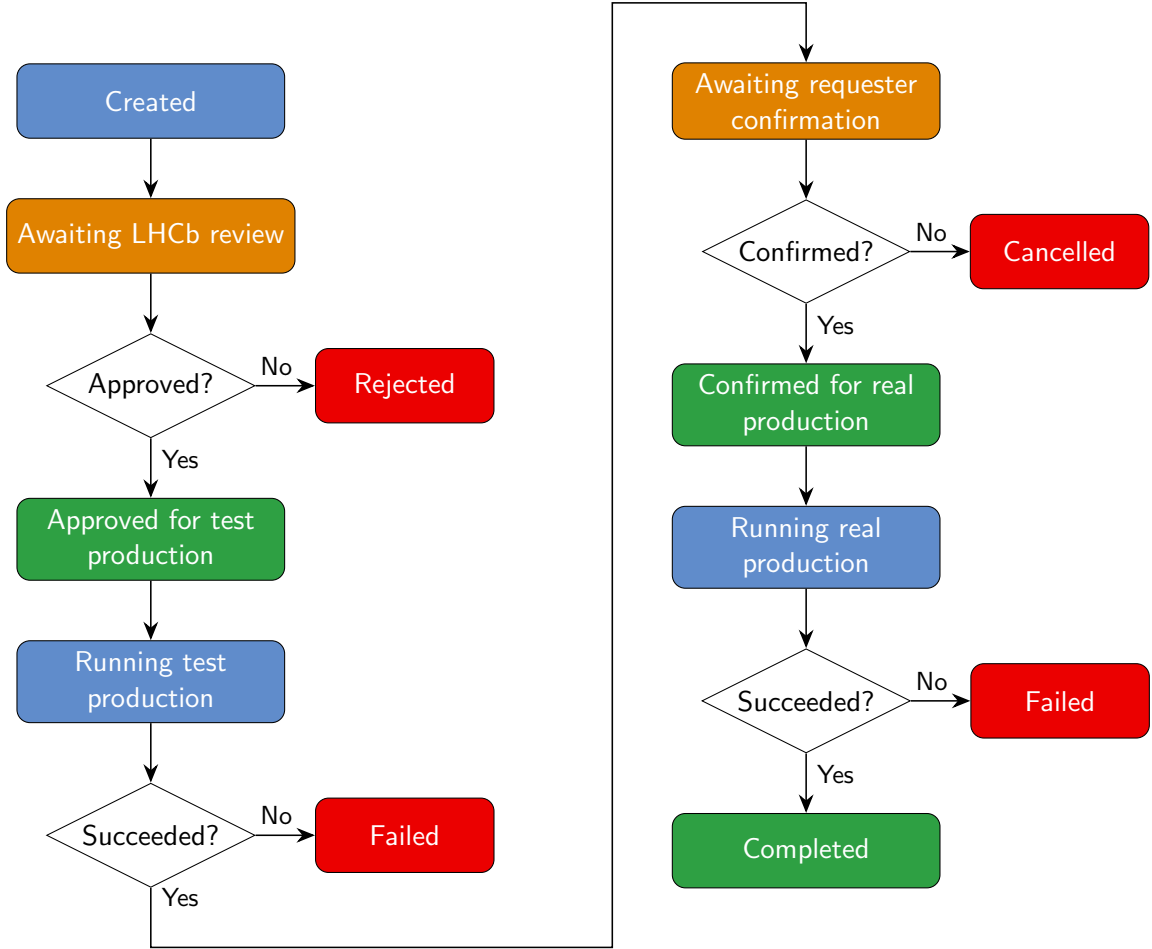

        Each request flows through a set of statuses modeled as GitLab issue labels as shown in Figure~\ref{fig:gitlab_labels}.
    
        A request may terminate for one of several reasons: failure during the running of the test/full production; cancellation by the requester after reviewing the test production results; or rejection by the \lhcb team after review if a problem with the configuration is spotted. The \lhcb team can approve/reject requests from GitLab by changing issue labels. Most of the other status changes are event-driven and handled automatically by the Ntupling Service using GitLab's API.
    
        The test production runs over a small subset of the selected data stream(s), and the resulting ntuples are returned to the user upon completion so that the user can inspect them to verify that the desired variables are available, and use them to set up and test their analysis workflow. After reviewing the test production results, the user can either:
    
        \begin{itemize}
            \item Proceed to triggering the full production; or
            \item Make the appropriate modifications, and submit a new request.
        \end{itemize}
        
        For the latter case, a feature to clone configurations of past requests can be used as a starting point. For the former, the LHCb Open Data team is informed and performs a final inspection before merging the configuration files into the Analysis Productions repository to trigger the full production. Upon completion of the full production, the ntuples are copied to a location in EOS which is exposed to the service backend. The user can then view and download the result files from the web interface. Users are updated by email upon request status changes and when the \lhcb team comments on their request(s).

    \subsection{Security Measures}
    
        Security measures have been put in place to ensure that no malicious code can be injected into the WLCG or the \lhcb file system through the service.
    
        %\subsubsection{Production file generation}
    
        Configuration of production jobs from the wizard is captured in static \texttt{YAML} files. The service never accepts executable code from the user. Rather, the backend itself generates the necessary Python production files from a request's \texttt{YAML} files, so the code that runs in Analysis Productions originates from the Ntupling Service and not from the user. The backend also ensures that only the expected file types are received, and that configuration files have not been modified to access unreleased data samples. Because the backend also writes files to shared storage using user-supplied filenames, these are also validated to prevent directory-traversal attacks.
    
        %\subsubsection{Authentication and authorization}
    
        User authentication is delegated to CERN's SSO proxy, which injects the verified user email into the \texttt{X-Forwarded-User} header~\footnote{\texttt{The X-Forwarded-User} request header allows the identification of the originating IP address of a client connecting to a web server through a proxy server.} read by the backend. Because the backend can only be reached by going through the SSO proxy, this header cannot be forged externally. In addition, operations on a request (commenting, canceling, approving results, etc.) are protected by a check comparing the SSO-asserted email against the requester's email in the GitLab issue, so users can only act on their own requests.

        %\subsubsection{Manual control}

        Finally, the \lhcb team retains manual control over much of the request workflow. No request is approved automatically: the team reviews each one by hand and controls when to run it on Analysis Productions, rather than being triggered by the service itself. While the safeguards above defend against known attack surfaces related to malformed/malicious configurations, this human review provides an extra layer of security, ensuring that nothing runs on the WLCG without a person having inspected and approved it.

\section{User Experience}
    \label{sec:user-experience}
    
    A core design principle of the \lhcb Ntupling Service is to provide an intuitive and guided user experience. In developing the web service, particular emphasis was placed on lowering the barrier to requesting \lhcb open data as much as possible. \Cref{sec:docu,sec:wizard} describe the measures taken to support this goal, including comprehensive documentation and in-application guidance that help users understand the data they are requesting. Sections~\ref{sec:communication} and~\ref{sec:usage} discuss the primary communication channel for user support, as well as initial feedback and usage statistics for the \lhcb Ntupling Service. Finally, \Cref{sec:story} summarizes different use cases encountered for the service so far, and feedback obtained from users.

    \subsection{Guiding a First-Time User}
    \label{sec:docu}
\begin{figure}[ht]
        \centering
        \includegraphics[width=0.95\linewidth]{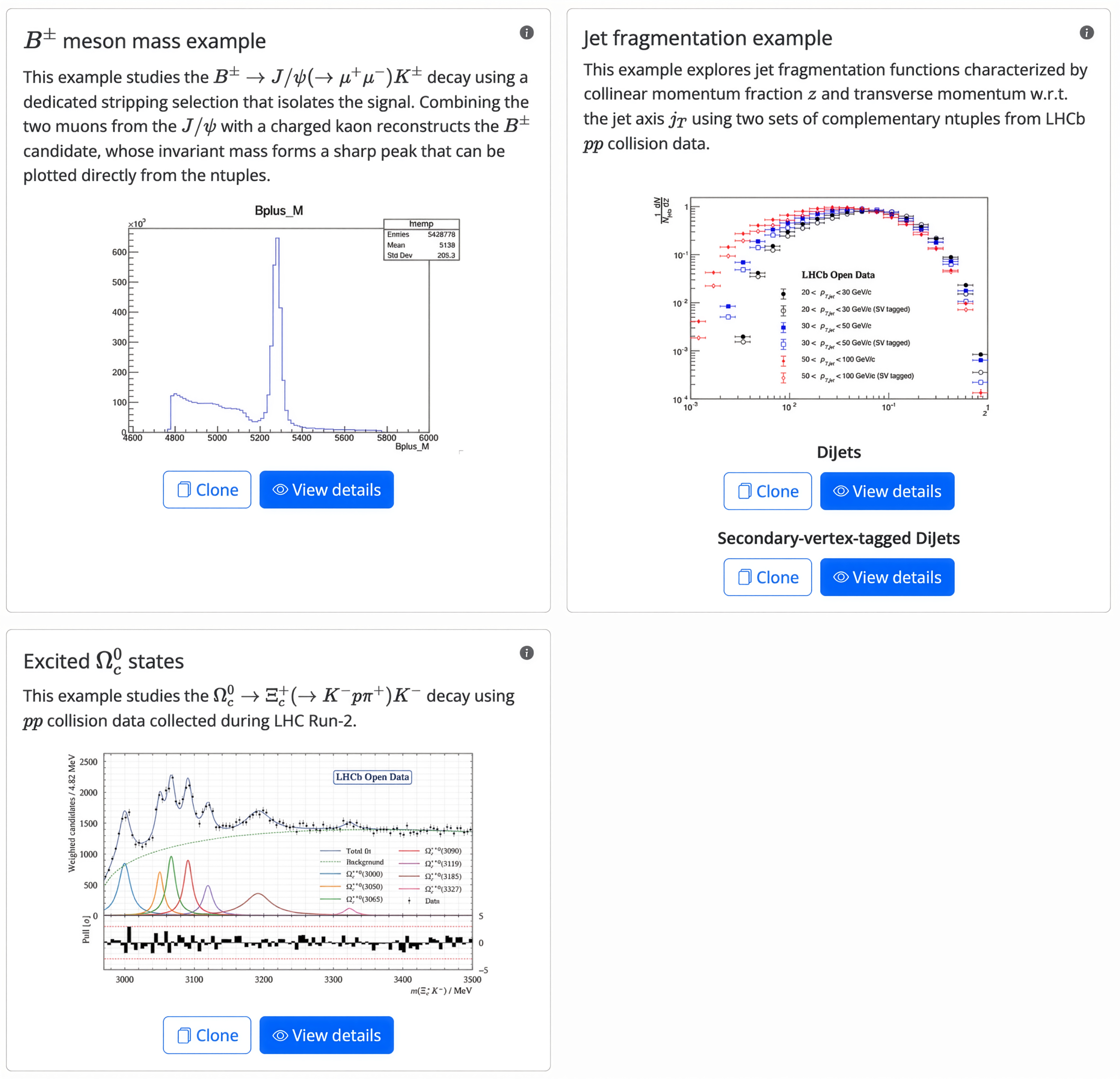}
        \caption{The gallery page for viewing example requests, working with preproduced ntuples, and / or cloning the corresponding request configuration as a template for a new request.}
        \label{fig:gallery}
    \end{figure}

\begin{figure}[htbp]
        \centering
        \includegraphics[scale=0.8]{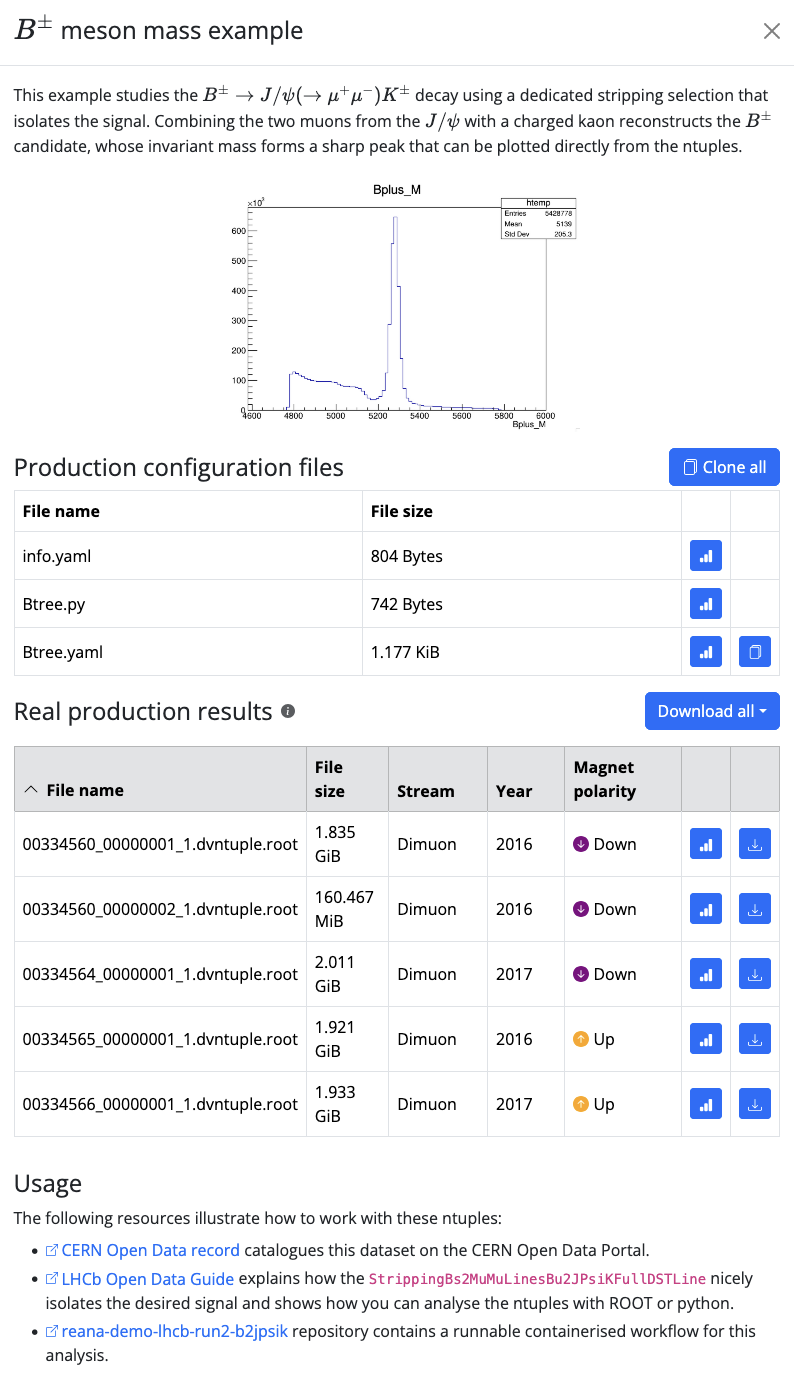}
        \caption{Expanded view of the \Bpm meson mass example using the \Bpm \to \jpsi \Kpm decay channel.}
        \label{fig:gallery-expanded}
    \end{figure}

        Upon first visiting the website, users are met with a welcome message, and the opportunity to visit a growing gallery of example requests with corresponding ntuples for inspiration, or to create their own request, as shown in~\Cref{fig:gallery}. An expanded view of the \Bpm meson mass gallery example is shown in \Cref{fig:gallery-expanded}. Gallery examples are also accompanied by open data records on the CERN Open Data Portal, with Digital Object Identifiers (DOIs), citation templates and a detailed usage guide~\cite{lhcb_opendata_guide} on how to analyze the example data including REANA~\cite{REANA} reproducible analysis workflows.

        The usage guide provides step-by-step instructions for using the web service, from request creation, the selection of \lhcb datasets, physics objects and output variables, to submission and progression tracking. It also explains common \lhcb nomenclature throughout the documentation. Additionally, the various pages encountered when configuring or viewing requests link directly to relevant sections of the guide. The usage guide also hosts analysis examples for the gallery requests, providing easy access to ROOT and Python code that demonstrates typical \lhcb workflows, from reading and filtering ntuples to performing downstream analysis. 
        
    \subsection{Creating a Custom LHCb Open Data Request}
        \label{sec:wizard}

\begin{figure}[htbp]
        \centering
            \centering
            \includegraphics[width=\linewidth]{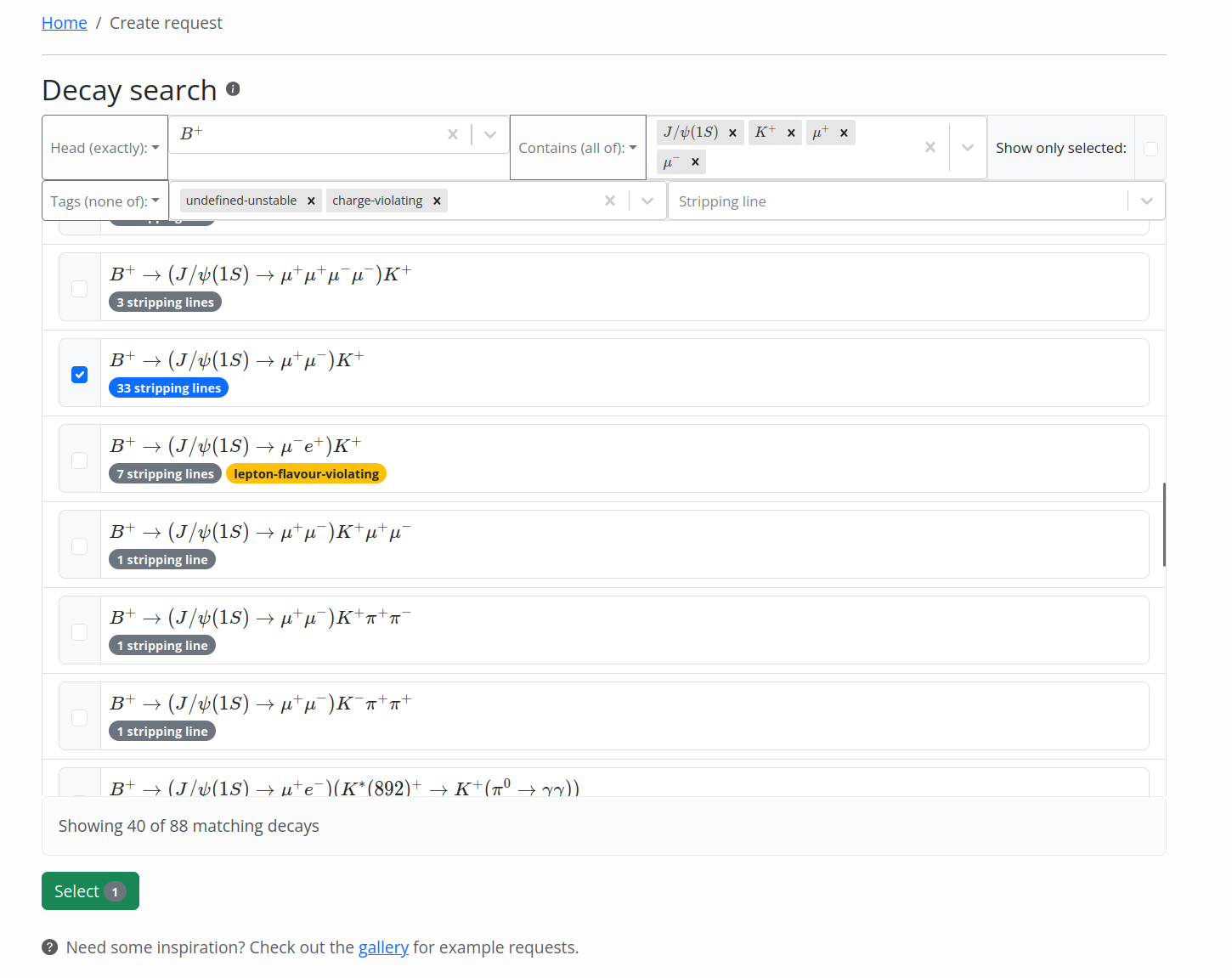}
            \caption{Physics object selection step of the request configuration process.}
            \label{fig:custom-request-decay}
        \end{figure}

\begin{figure}[htbp]
            \centering
            \includegraphics[width=\linewidth]{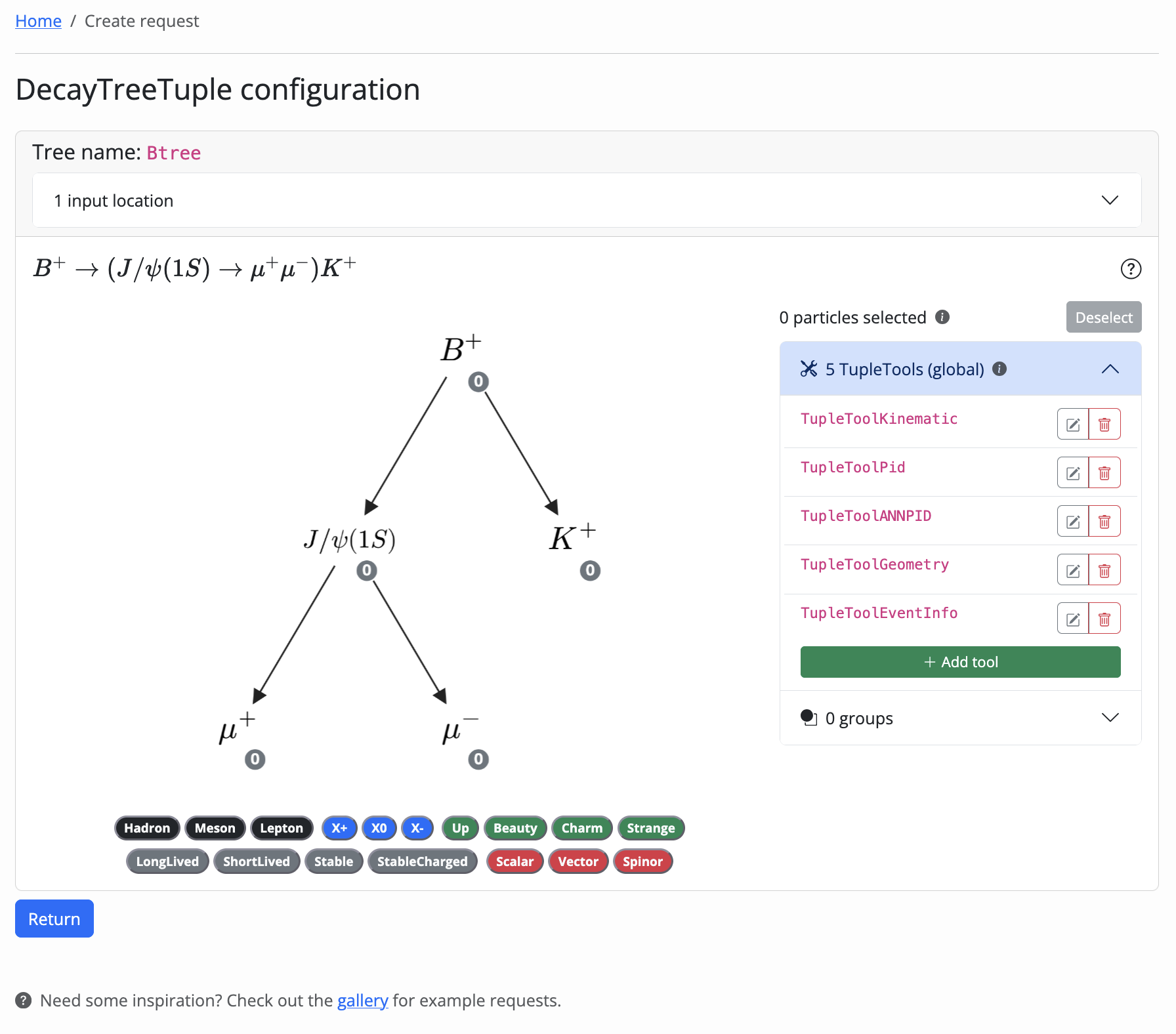}
            \caption{Custom configuration of columns to be added to the ntuple during the request configuration process.}
            \label{fig:custom-request-tupleconfig}
        \end{figure}

\begin{figure}[htbp]
            \centering
            \includegraphics[width=\linewidth]{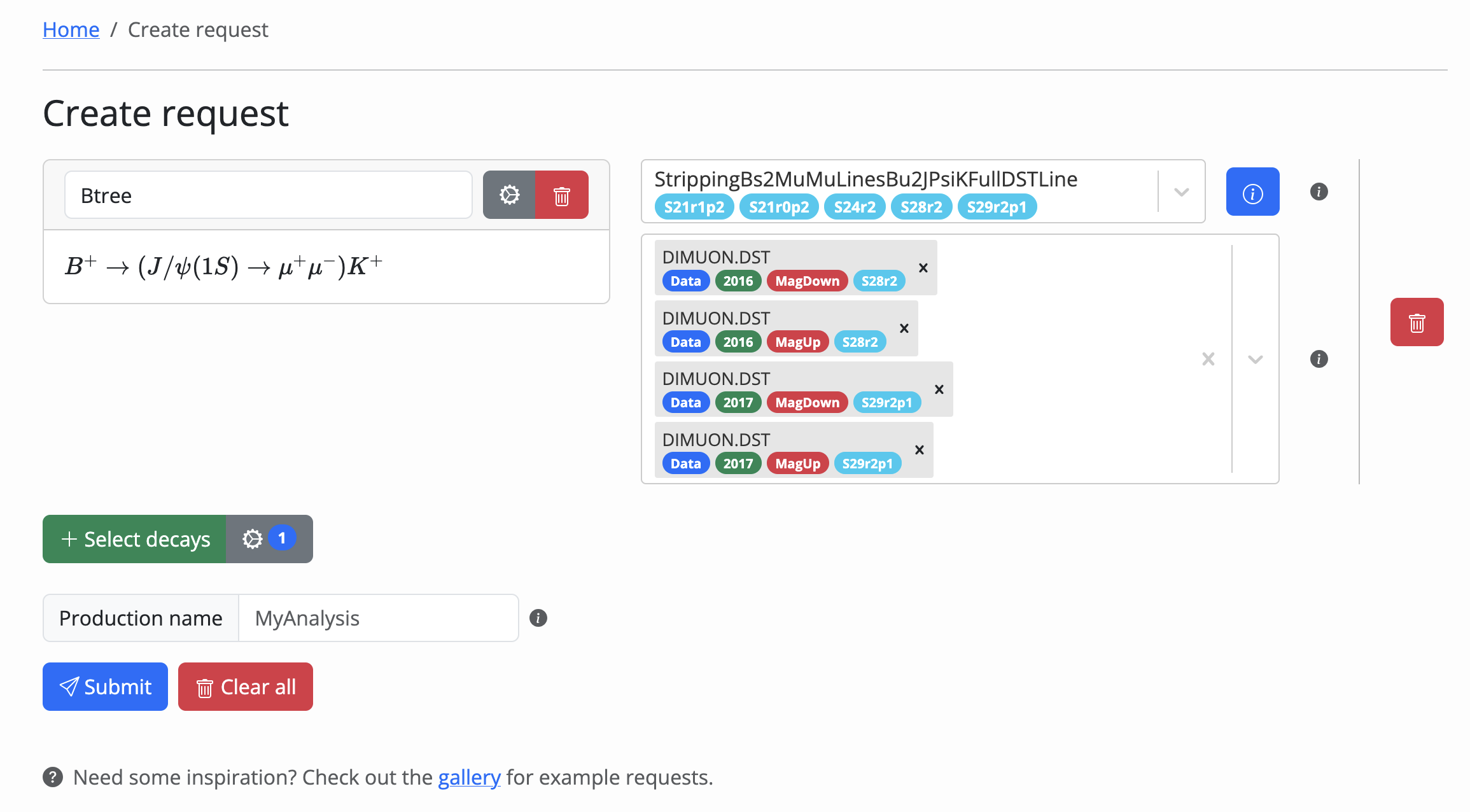}
            \caption{Specification of an LHCb dataset during the request configuration process.}
            \label{fig:custom-request-datasets}
    \end{figure}

        When preparing a new request, users first select the available physics objects, typically by choosing a physics decay process. At this stage, the search can already be constrained by selecting tags, particles contained in the decay, or so-called \textit{Stripping} selection lines~\cite{lhcb_stripping_paper}, which correspond to the output of the \lhcb preselection processing. This interface is shown in \Cref{fig:custom-request-decay}. As illustrated in \Cref{fig:custom-request-tupleconfig}, the configuration of the requested ntuple(s) is fully customizable and relies on \lhcb-internal tools to write the desired variables to the output ntuple(s). A set of default tools is provided to include commonly used variables, such as information about particle kinematics and geometry, and operational metadata from the LHC machine and the \lhcb experiment. Finally, users specify the data to be processed by \lhcb on the WLCG on their behalf. The available datasets are categorized into physically meaningful subsets of the collision data, referred to as \textit{Streams}, as well as by data-taking year and \lhcb magnet polarity. This step is shown in \Cref{fig:custom-request-datasets}. Detailed information regarding the request generation procedure can be found in Ref.~\cite{lhcb_ntuple_wizard_paper}.

    
    \subsection{Communication and Request Tracking}
    \label{sec:communication}
\begin{figure}[htbp]
        \centering
        \includegraphics[width=0.95\linewidth]{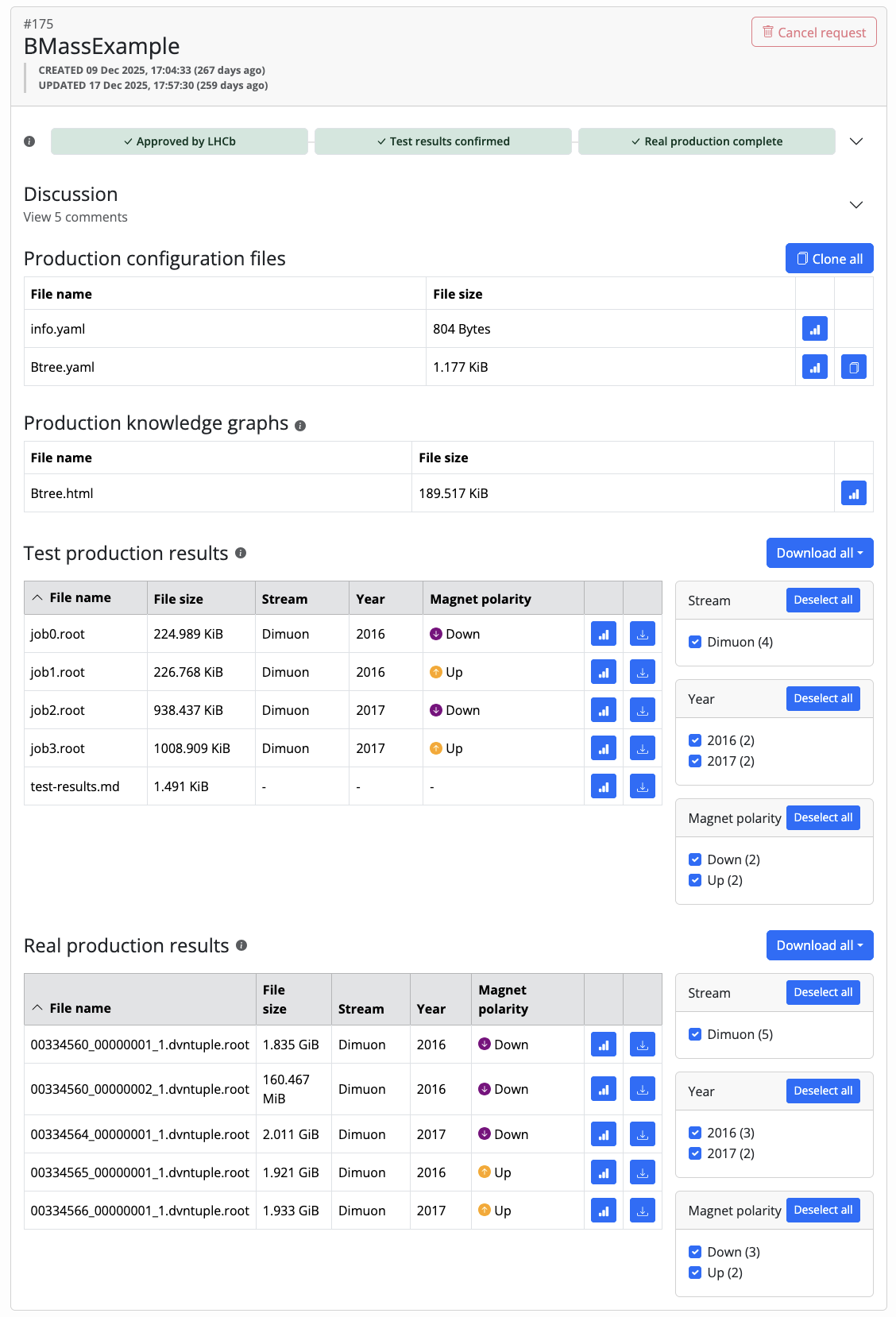}
        \caption{Example of the page users interact with to view their requests.}
        \label{fig:view_request}
    \end{figure}

\begin{figure}[htbp]
        \centering
        \includegraphics[width=0.95\linewidth]{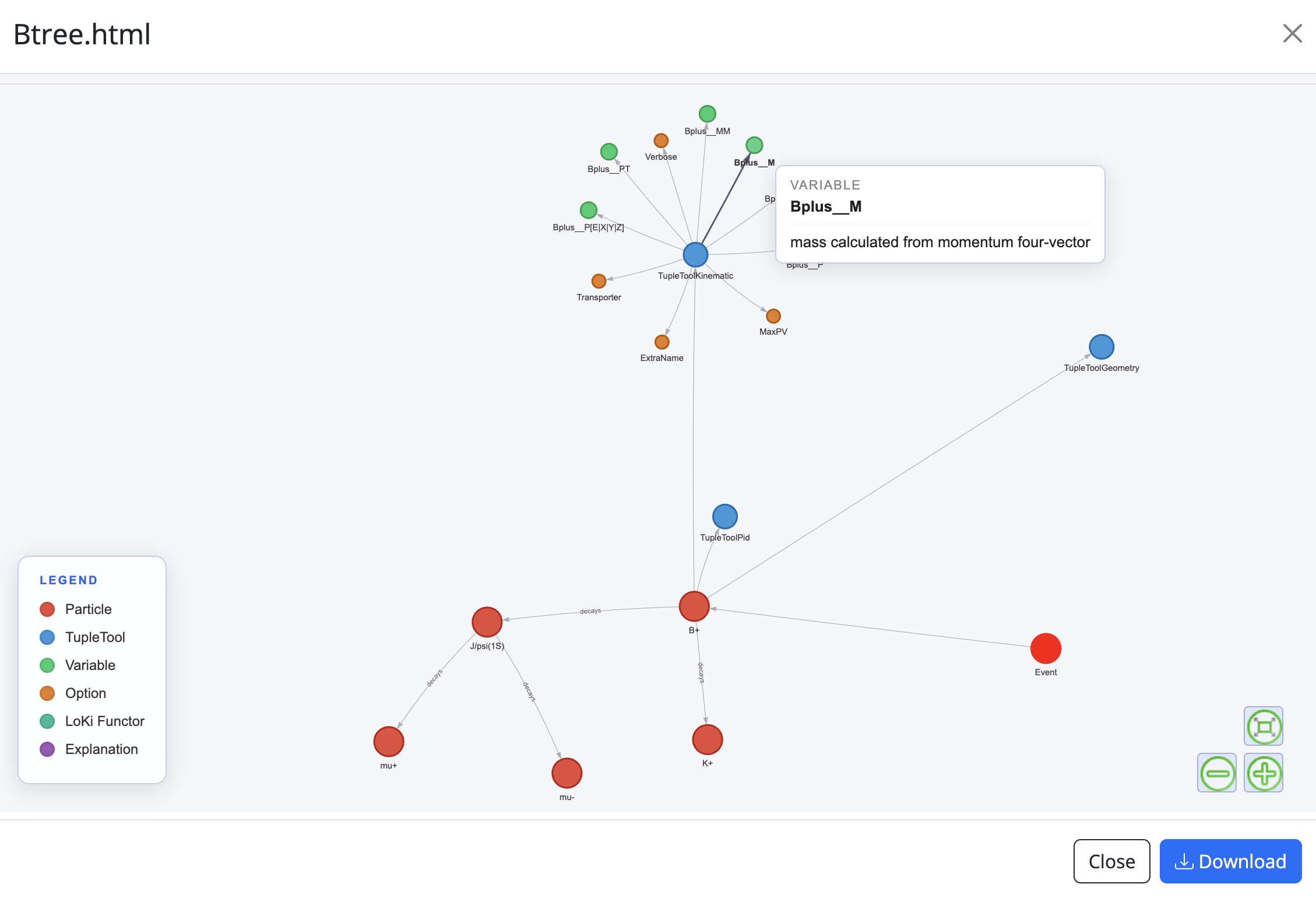}
        \caption{Example of the knowledge graph corresponding to Figure~\ref{fig:view_request}, allowing for visualization of available variables in the ntuple.}
        \label{fig:knowledge_graph}
    \end{figure}

        The service is an all-in-one hub for request submission, request tracking, and output retrieval. The default landing page displays a table with information about the user's active and past requests, including the creation date, production name, and request status. Clicking on a table entry routes to a dedicated request details page, where users can view the full status of their request, the corresponding production configuration files, output ntuples from both the test and full production stages, and a knowledge graph describing many of the variables saved to the ntuple. An example of the request details page can be seen in Figure~\ref{fig:view_request}, with the corresponding knowledge graph shown in Figure~\ref{fig:knowledge_graph}.
        
        The output files are categorized according to the datasets from which they were produced, allowing users to easily identify which data-taking year, magnet polarity, and stream a given ntuple corresponds to. From the request details page, users can inspect ROOT files directly in the browser using JSROOT~\cite{JSROOT}, displaying histograms of data variables. They can also retrieve the produced ntuples either through direct download from the web interface or via supported access protocols such as XRootD~\cite{XRootD}.
        
        The same page also provides a request-specific communication channel between the user and the \lhcb Open Data team. This allows users to raise questions or report concerns directly in the context of a given request, enabling the team to provide targeted, on-demand support at the request level. These messages are displayed in the request database, allowing the Open Data team to respond quickly and keep track of request-specific issues. In addition, users are notified by email for any communication from the \lhcb Open Data team, and about relevant status changes, for example when GitLab labels associated with their request are updated, ensuring that they remain informed throughout the production workflow.

    \subsection{Usage Uptake}
    \label{sec:usage}
    
    Since the public release of the \lhcb Ntupling Service on February 22, 2026, the number of submitted requests has steadily increased. Although the service has only been publicly available for about six months at the time of writing, we have accumulated roughly 100 user requests. 
    
    The size of the produced outputs varies substantially between requests, depending on the selected datasets, data-taking years, and requested analysis variables. Small requests may produce output ntuples of only a few hundred MB, while larger requests can result in several hundred GB of output data. For the first 100 requests, the total produced output amounted to approximately 10\,TB, corresponding to an average output size of about 100\,GB per request.

    \subsection{User Feedback}
    \label{sec:story}
    
    As part of the request submission procedure, users are asked to briefly describe the intended use of the requested \lhcb open data. These descriptions provide useful insight into motivations for accessing ntuples through the \lhcb Ntupling Service -- they are most relevant for data popularity studies.
    
    The submitted use cases cover a broad range of applications. Some users request data for physics studies, including CP-asymmetry measurements, searches for exotic hadrons, and studies of detector resolution effects. Others use the service to investigate statistical and computational methods, such as fitting algorithms or resolution models, in a realistic HEP environment.
    
    The service is also used in educational contexts. Examples include high-school-level outreach activities, university teaching, and self-guided exercises for learning common analysis tools such as ROOT. In addition, the service provides a lightweight entry point to \lhcb data even within the collaboration, for example when training new students before they interact with the full (complexity of the) internal data-processing infrastructure.
    
    The request-specific communication channel has proven useful for understanding user needs and for providing targeted support. Feedback received through these conversations is generally very positive.
    The ability to ask questions directly in the context of a given request is particularly appreciated. The repeated use of the service by several users after an initial request further suggests that the workflow is useful beyond isolated test cases. At the same time, this feedback is valuable for the development of the service itself, since recurring questions and difficulties help identify where the interface, documentation, or analysis examples can be improved.
    
\section{CERN Open Data Portal Integration}
\label{sec:ODP-integration}
    The CERN Open Data Portal and the \lhcb Ntupling Service operate as separate entities. The CERN Open Data Portal is designed to permanently host static datasets, and it serves as a central hub for the LHC experiments and beyond to host open data. The needs of the \lhcb Ntupling Service are fundamentally different, as ntuple creation is done dynamically, and resulting ntuples are not stored permanently unless they are promoted to permanent records by the \lhcb Open Data team. However, there are still opportunities for integration between the CERN Open Data Portal and the \lhcb Ntupling Service. For example, Section~\ref{sec:record_promotion} discusses the feature of promotion of service requests to permanent records on the CERN ODP, helping to facilitate FAIR principles~\cite{FAIR} for data output from the service. In turn, the \lhcb Ntupling Service is advertised directly on the CERN Open Data Portal on both the ``Getting started with LHCb Open Data'' and ``About LHCb'' pages to promote findability and accessibility of the web service. 

    \subsection{Promotion to Permanent Records}
    \label{sec:record_promotion}
    
\begin{figure}[htbp]
            \centering
            \includegraphics[width=0.98\textwidth]{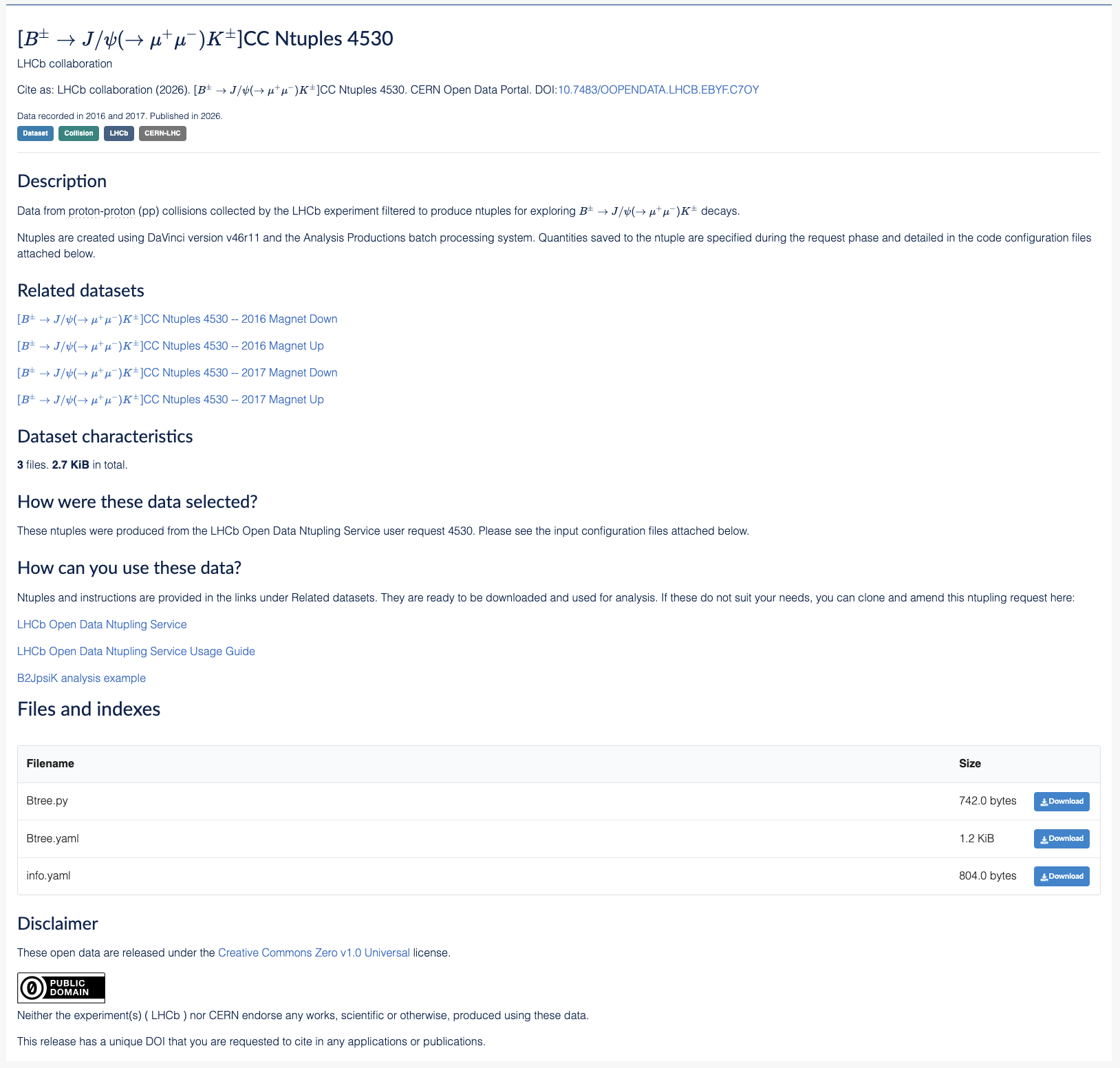}
            \caption{Example of a record on the CERN Open Data Portal resulting from a request from the \lhcb Ntupling Service.}
            \label{fig:ODP-record}
        \end{figure}

    Requests on the service can be promoted to permanent records on the CERN Open Data Portal; a DOI is attributed to the record and corresponding ntuples, and instructions for citation of the data are provided. A procedure is in development to aggregate the necessary metadata and produce the Open Data Portal record in the appropriate \texttt{JSON} schema during the request life cycle. Users will be able to request promotion of their ntuples to a permanent dataset on the CERN Open Data Portal, which will ultimately need to be approved by the \lhcb Open Data team. A permanent record for the \Bpm meson mass example discussed in Section~\ref{sec:user-experience} is shown in \Cref{fig:ODP-record}. As such records accumulate, the gallery of examples advertised on the service will grow, which will directly cross link the CERN Open Data Portal records on the service and provide the option to work with these ntuples rather than producing new ones. This will lead to a favorable trade off of the compute on the WLCG for the relatively light footprint of storing ntuples on the CERN Open Data Portal. 
    
    \subsection{Storage Considerations}
        As mentioned previously, ntuples produced by the \lhcb Ntupling Service are not stored permanently by default. Instead, users are prompted to download their resulting ntuples onto local data volumes. The time window before deletion of the produced ntuples on the service side is still being decided, and will be informed based on average accumulation of data volume. On the other hand, requests promoted to permanent records on the Open Data Portal will have ntuples stored indefinitely for future use. Both the temporary and permanent data stored from the service must be taken into account relative to the total data volume allocated to the \lhcb open data project. At the time of writing, less than 10\,TB have been accumulated from \lhcb Ntupling Service output, which is only a small fraction of the data volume allocated to \lhcb for open data activities. The current rate of data accumulation should give ample time to develop an appropriate solution for temporary storage cleanup, including informing users well in advance before any data deletion.

\section{Future Outlook}
\label{sec:future-outlook}
    The \lhcb collaboration has now provided open access to ${\sim}100\%$ of the Level-3 proton-proton ($pp$) collision data collected during LHC's Run 1 (2011-2012) and over $50\%$ of the Level-3 $pp$ collision data collected during Run 2 (2015-2018), in accordance with \lhcb's implementation of the CERN Open Data Policy~\cite{CERN_Open_Data_Policy}. This outlines \lhcb's commitment to making its data publicly accessible: 50\% of the data from a given data-taking campaign is to be released five years after its completion, followed by the full dataset after ten years. Level-3 data in this context refers to reconstructed data suitable for research purposes, as defined by the Data Preservation in High Energy Physics (DPHEP) working group~\cite{DPHEP_2012}. The current implementation in the \lhcb Ntupling Service includes the 2011 and 2012 data for Run 1, and the 2015-2017 data for Run 2, excluding the 2018 data. This accounts for roughly all of the integrated luminosity for Run 1 and over half of the integrated luminosity for Run 2. In 2028 it will be 10 years out from Run 2 data taking, prompting the need to provide access to the 2018 data in the service at this time. This will be a simple operation by design, leading to no foreseen delays in providing open access to the full Run 2 \lhcb dataset by 2028. 

    In a similar manner, releases of data collected in other collision systems (such as $p$Pb, Pb$p$, and PbPb in collider mode, and collisions of protons or lead on noble gas species in fixed target mode with the System for Measuring Overlap with Gas (SMOG))~\cite{heavy_ion, SMOG} are achievable with relative ease. Future developments will proceed based on the resources available to the \lhcb Open Data project. 

    No simulation samples are available for request with the LHCb Ntupling Service at the time of writing. However, the LHCb simulation stack is open source, and a documented procedure for generating Monte Carlo (MC) and running it through the LHCb detector simulation as an external user is in development. This would require users to provide their own computing resources in order to generate and process the simulation samples. Other means for providing access to previously prepared simulation samples where only the ntupling step is required are being investigated in parallel. In a similar manner, there is an effort to provide access to calibration samples and internally available tools for calculating track, PID, and trigger efficiencies, but the timescale for achieving this is not yet clear. While this currently presents limitations for publication-grade analyses, these limitations can be overcome with the aforementioned future developments, and the LHCb Open Data team will do what they can to help users in overcoming these hurdles in the meantime. All developments in this direction will proceed according to resources available to the \lhcb Open Data project. 

    Releases for Run 3 and beyond will take considerable effort, as the infrastructure for ntupling at \lhcb has been completely reworked (and improved)~\cite{FunTuple}. This means that the configuration files derived from the wizard and corresponding internal parsing package described in Ref.~\cite{lhcb_ntuple_wizard_paper} would need a major overhaul and separately managed logic for legacy (Run 1 and Run 2) productions and recent and future (Run 3 and beyond) productions.
    Such an endeavor remains to be discussed within the \lhcb collaboration.
    
\section{Summary}
\label{sec:summary}
    The web service presented in this paper, the \lhcb Ntupling Service, constitutes a new paradigm for providing open access to data collected in high energy physics experiments. In conjunction with the \lhcb Ntuple Wizard, this innovative framework provides access to both Run 1, and for the first time, Run 2 $pp$ data collected by the \lhcb experiment, amounting to over 4~PB of data to explore. Future releases of 2018 Run 2 data, as well as Run 1 and 2 data with proton-ion or ion-ion collisions can be accomplished with relative ease by design, though releases for Run 3 and beyond will take considerable effort given the changes to the ntuple production software at \lhcb. The service provides flexibility for deep exploration of the collected data while reducing the barrier to entry for new users and maintaining a significantly smaller footprint in terms of data volumes relative to historical release mechanisms. It has a robust architecture connecting the \lhcb Ntuple Wizard for specification of request configuration, with the Analysis Productions batch processing system for running production jobs on the WLCG. It also provides a single entry point for users to submit ntuple production jobs, track their requests, communicate with the \lhcb Open Data team, and download their resulting output files. Substantial documentation and usage hints are provided on the \lhcb Open Data Guide, and are embedded directly into the web service. User requests can be promoted to permanent records on the CERN Open Data Portal upon approval from the LHCb Open Data team, in which case the resulting ntuples are stored permanently with citable DOIs.
    
    The \lhcb Ntupling Service was officially released in the first quarter of 2026 and has since seen sustained usage with very positive feedback. A wealth of use cases has already been observed, including research into exotic hadrons, development of fitting algorithms, physics education, and much more. This is a significant milestone for both the \lhcb experiment and open science in general, allowing for exploration of the unique data collected by the \lhcb experiment without the need to learn any experiment-specific software.

%% file: acknowledgements.tex
\section*{Acknowledgements}

We thank our colleagues of the LHCb Data Processing \& Analysis (DPA) project for their incredibly valuable discussions and enthusiasm. We thank the LHCb Computing and DPA projects for providing the necessary resources for the LHCb Ntupling Service, through the usage of Analysis Productions ran on the WLCG facilities. We also thank the Open Science team at CERN's Department of Information Technology for their extensive support of the LHCb Ntupling Service and for providing storage space for the resulting data outputs. C.A.A. and D.S.F. acknowledge support from the National Science Foundation under Award Number 2309922. We further acknowledge support from the national German agencies BMFTR and PUNCH4NFDI (DFG).

%% file: LHCb/LHCb-TDR.bib
@lhcbreport{LHCb-TDR-017,
      collaboration = "LHCb collaboration",
      title         = "{LHCb Upgrade Software and Computing}",
      institution   = "CERN",
      address       = "Geneva",
      number        = "{CERN-LHCC-2018-007}",
      year          = "2018",
}

@lhcbreport{LHCb-TDR-011,
      collaboration = "LHCb collaboration",
      title         = "{LHCb computing: Technical Design Report}",
      institution   = "CERN",
      address       = "Geneva",
      number        = "{CERN-LHCC-2005-019}",
      year          = "2005",
}


%% file: LHCb/standard.bib
@article{Root,
    author         = "Brun, R. and Rademakers, F.",
    title          = "{ROOT -- An object oriented data analysis framework}",
    booktitle      = "{New computing techniques in physics research V. Proceedings, 5th International Workshop, AIHENP '96, Lausanne, Switzerland, September 2-6, 1996}",
    journal        = "Nucl. Instrum. Meth.",
    volume         = "A389",
    year           = "1997",
    pages          = "81-86",
    doi            = "10.1016/S0168-9002(97)00048-X",
    SLACcitation   = "%%CITATION = NUIMA,A389,81;%%"
}

@article{FunTuple,
      author         = "Mathad, Abhijit and  Ferrillo, Martina and  Barr{\'e}, Sacha and  Koppenburg, Patrick and  Owen, Patrick and  Raven, Gerhard and  Rodrigues, Eduardo and  Serra, Nicola",
      title              = "{FunTuple: A new N-tuple component for offline data processing at the LHCb experiment}",     
      journal        = "Comput. Softw. Big Sci. ",
      volume       = "8",
      year           = "2024",
      pages        = "6",
      archivePrefix  = "arXiv",
      primaryClass   = "physics.data-an",
      eprint   = "2310.02433",
      doi              = "10.1007/s41781-024-00116-1",
}

@article{Stripping,
      author        = "Grieser, Nathan and Rodrigues, Eduardo and Sahoo, Niladri and Sheng, 
                       Shuqi and Skidmore, Nicole and Smith, Mark",
      title         = "{The LHCb stripping project: Sustainable legacy data processing for high-energy physics}",
      eprint        = "2509.05294",
      archivePrefix = "arXiv",
      primaryClass  = "hep-ex",
      year          = "2025",
      journal       = "Comput. Softw. Big Sci.",
      volume        = "9",
      pages         = "21",
      doi           = "10.1007/s41781-025-00151-6"
}


%% file: main.bib
@misc{lhcb_ntupling_service,
  collaboration= "{LHCb}",
  title        = "{LHCb Ntupling Service}",
  publisher    = "",
  year         = "2026",
  howpublished = "\url{https://opendata-lhcb-ntupling-service.app.cern.ch/}"
}

@article{lhcb_ntuple_wizard_paper,
    author = "Aidala, Christine A. and Burr, Christopher and Cattaneo, Marco and Fitzgerald, Dillon S. and Morris, Adam and Neubert, Sebastian and Tropmann, Donijor",
    title = "{Ntuple Wizard: An Application to Access Large-Scale Open Data from LHCb}",
    eprint = "2302.14235",
    archivePrefix = "arXiv",
    primaryClass = "hep-ex",
    doi = "10.1007/s41781-023-00099-5",
    journal = "Comput. Softw. Big Sci.",
    volume = "7",
    number = "1",
    pages = "6",
    year = "2023"
}

@article{lhcb_analysis_productions_paper,
    author = "Abdelmotteleb, Ahmed and others",
    title = "{The LHCb Sprucing and Analysis Productions}",
    eprint = "2506.20309",
    archivePrefix = "arXiv",
    primaryClass = "hep-ex",
    doi = "10.1007/s41781-025-00144-5",
    journal = "Comput. Softw. Big Sci.",
    volume = "9",
    number = "1",
    pages = "15",
    year = "2025"
}

@article{lhcb_stripping_paper,
    author = "Grieser, Nathan Allen and Rodrigues, Eduardo and Sahoo, Niladri and Sheng, Shuqi and Skidmore, Nicole and Smith, Mark",
    title = "{The LHCb Stripping Project: Sustainable Legacy Data Processing for High-Energy Physics}",
    eprint = "2509.05294",
    archivePrefix = "arXiv",
    primaryClass = "hep-ex",
    doi = "10.1007/s41781-025-00151-6",
    journal = "Comput. Softw. Big Sci.",
    volume = "9",
    number = "1",
    pages = "21",
    year = "2025"
}

@misc{lhcb_ntuple_wizard_npm,
  author       = "{LHCb collaboration}",
  title        = "{LHCb Ntuple Wizard}",
  publisher    = "node package manager",
  year         = "2026",
  howpublished = "\url{https://www.npmjs.com/package/lhcb-ntuple-wizard}"
}

@misc{lhcb_ntuple_wizard_git,
  author       = "{LHCb collaboration}",
  title        = "{LHCb Ntuple Wizard}",
  publisher    = "GitLab",
  year         = "2026",
  howpublished = "\url{https://gitlab.cern.ch/lhcb-dpa/wp6-analysis-preservation-and-open-data/lhcb-ntuple-wizard-frontend}"
}

@misc{gitlab,
  author       = {{GitLab Inc\period}},
  title        = "{GitLab}",
  publisher    = "GitLab",
  howpublished = "\url{https://about.gitlab.com/}"
}

@misc{npm,
  author       = "{npm}",
  title        = "{npm}",
  publisher    = "npm",
  year         = "",
  howpublished = "\url{https://www.npmjs.com/}"
}

@misc{lhcb_opendata_guide,
  author       = "{LHCb collaboration}",
  title        = "{LHCb Open Data Guide}",
  publisher    = "",
  year         = "2026",
  howpublished = "\url{https://lhcb-opendata-guide.web.cern.ch/}"
}

@misc{CERN_ODP,
  author       = "{CERN}",
  title        = "{CERN Open Data Portal}",
  publisher    = "",
  year         = "",
  howpublished = "\url{https://opendata.cern.ch/}"
}

@article{FAIR,
    author = "Wilkinson, Mark D. and others",
    title = "{The FAIR Guiding Principles for scientific data management and stewardship}",
    doi = "10.1038/sdata.2016.18",
    journal = "Sci. Data",
    volume = "3",
    number = "1",
    pages = "160018",
    year = "2016"
}

@article{service_proceedings,
    author = "Aidala, Christine and others",
    title = "{LHCb Open Data Ntupling Service: On-demand production and publishing of custom LHCb Open Data}",
    eprint = "2504.00610",
    archivePrefix = "arXiv",
    primaryClass = "hep-ex",
    doi = "10.1051/epjconf/202533701231",
    journal = "EPJ Web Conf.",
    volume = "337",
    pages = "01231",
    year = "2025"
}

@article{REANA,
    author = "{\v{S}}imko, Tibor and Heinrich, Lukas and Hirvonsalo, Harri and Kousidis, Dinos and Rodr{\'\i}guez, Diego",
    editor = "Forti, A. and Betev, L. and Litmaath, M. and Smirnova, O. and Hristov, P.",
    title = "{REANA: A System for Reusable Research Data Analyses}",
    reportNumber = "CERN-IT-2018-003",
    doi = "10.1051/epjconf/201921406034",
    journal = "EPJ Web Conf.",
    volume = "214",
    pages = "06034",
    year = "2019"
}

@article{Antcheva:2009zz,
    author = "Antcheva, I. and others",
    title = "{ROOT: A C++ framework for petabyte data storage, statistical analysis and visualization}",
    eprint = "1508.07749",
    archivePrefix = "arXiv",
    primaryClass = "physics.data-an",
    reportNumber = "FERMILAB-PUB-09-661-CD",
    doi = "10.1016/j.cpc.2009.08.005",
    journal = "Comput. Phys. Commun.",
    volume = "180",
    pages = "2499--2512",
    year = "2009"
}

@article{XRootD,
    author = "Amadio, Guilherme and others",
    title = "{XRootD Client: a robust technology for LHC Run-3 and beyond}",
    doi = "10.1051/epjconf/202429501056",
    journal = "EPJ Web Conf.",
    volume = "295",
    pages = "01056",
    year = "2024"
}

@article{JSROOT,
    author = "Bellenot, Bertrand and Linev, Sergey",
    title = "{JavaScript ROOT}",
    doi = "10.1088/1742-6596/664/6/062033",
    journal = "J. Phys. Conf. Ser.",
    volume = "664",
    pages = "062033",
    year = "2015"
}

@article{using-CMS-OD,
    author = "Lassila-Perini, Kati and Lange, Clemens and Jarrin, Edgar Carrera and Bellis, Matthew",
    title = "{Using CMS Open Data in research {\textendash} challenges and directions}",
    eprint = "2106.05726",
    archivePrefix = "arXiv",
    primaryClass = "hep-ex",
    doi = "10.1051/epjconf/202125101004",
    journal = "EPJ Web Conf.",
    volume = "251",
    pages = "01004",
    year = "2021"
}

@article{NANOAOD,
    author = {Ehat{\"a}ht, Karl},
    editor = "Doglioni, C. and Kim, D. and Stewart, G. A. and Silvestris, L. and Jackson, P. and Kamleh, W.",
    collaboration = "CMS",
    title = "{NANOAOD: a new compact event data format in CMS}",
    reportNumber = "CMS-CR-2020-048",
    doi = "10.1051/epjconf/202024506002",
    journal = "EPJ Web Conf.",
    volume = "245",
    pages = "06002",
    year = "2020"
}

@article{PHYSLITE,
	author = {{Schaarschmidt, Jana} and {Catmore, James} and {Elmsheuser, Johannes} and {Heinrich, Lukas} and {Krumnack, Nils} and {Mete, Serhan} and {Ozturk, Nurcan}},
	title = "{PHYSLITE - A new reduced common data format for ATLAS}",
	DOI= "10.1051/epjconf/202429506017",
	url= "https://doi.org/10.1051/epjconf/202429506017",
	journal = {EPJ Web of Conf.},
	year = 2024,
	volume = 295,
	pages = "06017",
}

@article{ATLAS_OD_release,
    author = "Albornoz, Mariana Vivas",
    collaboration = "ATLAS",
    title = "{The First Release of ATLAS Open Data for Research}",
    doi = "10.22323/1.476.1172",
    journal = "PoS",
    volume = "ICHEP2024",
    pages = "1172",
    year = "2025"
}

@misc{DaVinci_git,
  collaboration       = "LHCb",
  title        = "{DaVinci}",
  publisher    = "GitLab",
  year         = "",
  howpublished = "\url{https://gitlab.cern.ch/lhcb/DaVinci
}"
}

@techreport{ATLAS_computing_TDR,
    author = "Duckeck, G. and Barberis, D. and Hawkings, R. and Jones, R. and McCubbin, N. and Poulard, G. and Quarrie, D. and Wenaus, T. and Obreshkov, E.",
    collaboration = "ATLAS",
    institution   = "CERN",
    title = "{ATLAS computing: Technical Design Report}",
    number = "CERN-LHCC-2005-022, ATLAS-TDR-017",
    month = "6",
    year = "2005",
    url = "https://cds.cern.ch/record/837738",
}

@article{ATLAS_CSW_Run3,
    author = "Aad, G. and others",
    collaboration = "ATLAS",
    title = "{Software and computing for Run 3 of the ATLAS experiment at the LHC}",
    eprint = "2404.06335",
    archivePrefix = "arXiv",
    primaryClass = "hep-ex",
    reportNumber = "CERN-EP-2024-100",
    doi = "10.1140/epjc/s10052-024-13701-w",
    journal = "Eur. Phys. J. C",
    volume = "85",
    number = "3",
    pages = "234",
    year = "2025"
}

@techreport{CMS_computing_TDR,
      author        = "Bayatyan, G L and Della Negra, Michel and Foà, and
                       Hervé, A and Petrilli, Achille",
    collaboration = "CMS",
    institution   = "CERN",
    title = "{CMS computing: Technical Design Report}",
    number = "CERN-LHCC-2005-023, CMS-TDR-7",
    month = "6",
    year = "2005",
    url = "https://cds.cern.ch/record/838359",
}

@article{Mini_AOD,
    author = "Petrucciani, Giovanni and Rizzi, Andrea and Vuosalo, Carl",
    collaboration = "CMS",
    title = "{Mini-AOD: A New Analysis Data Format for CMS}",
    eprint = "1702.04685",
    archivePrefix = "arXiv",
    primaryClass = "physics.ins-det",
    reportNumber = "CMS-CR-2015-052",
    doi = "10.1088/1742-6596/664/7/072052",
    journal = "J. Phys. Conf. Ser.",
    volume = "664",
    number = "7",
    pages = "7",
    year = "2015"
}

@techreport{ALICE_computing_TDR,
      author        = "Cortese, P and Carminati, Federico and Fabjan, Christian
                       Wolfgang and Riccati, Lodovico and de Groot, Hans",
    collaboration = "ALICE",
    institution   = "CERN",
    title = "{ALICE computing: Technical Design Report}",
    number = "CERN-LHCC-2005-018, ALICE-TDR-12",
    month = "6",
    year = "2005",
    url ="https://cds.cern.ch/record/832753",
}

@techreport{ALICE_AO2D_TDR,
    author = "Buncic, P. and Krzewicki, M. and Vande Vyvre, P.",
    collaboration = "ALICE",
    institution   = "CERN",
    title = "{Technical Design Report for the Upgrade of the Online-Offline Computing System}",
    number = "CERN-LHCC-2015-006, ALICE-TDR-019",
    month = "4",
    year = "2015",
    url = "https://cds.cern.ch/record/2011297",
}

@techreport{CERN_Open_Data_Policy,
      title         = "{CERN Open Data Policy for the LHC Experiments}",
      institution   = "CERN",
      number  = "CERN-OPEN-2020-013",
      address       = "Geneva",
      year          = "2020",
      url           = "https://cds.cern.ch/record/2745133",
      doi           = "10.17181/CERN.QXNK.8L2G",
}

@article{DPHEP_2012,
    author = "Akopov, Zaven and others",
    collaboration = "DPHEP Study Group",
    title = "{Status Report of the DPHEP Study Group: Towards a Global Effort for Sustainable Data Preservation in High Energy Physics}",
    eprint = "1205.4667",
    archivePrefix = "arXiv",
    primaryClass = "hep-ex",
    reportNumber = "DPHEP-2012-001, FERMILAB-PUB-12-878-PPD",
    month = "5",
    year = "2012"
}

@article{nginx,
author = {Reese, Will},
year = {2008},
month = {01},
pages = {2},
title = {Nginx: The high-performance web server and reverse proxy},
volume = {2008},
journal = {Linux Journal}
}

@article{EOS,
    author = "Peters, AJ. and Sindrilaru, EA. and Adde, G.",
    title = "{EOS as the present and future solution for data storage at CERN}",
    doi = "10.1088/1742-6596/664/4/042042",
    journal = "J. Phys. Conf. Ser.",
    volume = "664",
    number = "4",
    pages = "042042",
    year = "2015"
}

@article{CephFS,
      author        = "Peters, Andreas J and van der Ster, Daniel C",
      title         = "{Evaluating CephFS Performance vs. Cost on High-Density
                       Commodity Disk Servers}",
      journal       = "Comput. Softw. Big Sci.",
      volume        = "5",
      number        = "1",
      pages         = "25",
      year          = "2021",
      url           = "https://cds.cern.ch/record/2790637",
      doi           = "10.1007/s41781-021-00071-1",
}

@misc{reactjs,
  author       = {{Meta Open Source}},
  title        = "{React: A JavaScript library for building user interfaces}",
  howpublished = {\url{https://react.dev/}},
}

@article{golang2022,
  author    = {Cox, Russ and Griesemer, Robert and Pike, Rob},
  title     = "{The Go Programming Language and Environment}",
  journal   = {Communications of the ACM},
  volume    = {65},
  number    = {5},
  pages     = {70--78},
  year      = {2022},
  month     = {may},
  publisher = {ACM},
  doi       = {10.1145/3488716}
}

@lhcbreport{SMOG,
    author = "Bursche, Albert and others",
    title = "{Physics opportunities with the fixed-target program of the LHCb experiment using an unpolarized gas target}",
    number = "LHCb-PUB-2018-015",
    year = "2018"
}

@lhcbreport{heavy_ion,
      author        = "Graziani, Giacomo and Massacrier, Laure Marie and Robbe,
                       Patrick and Schmelling, Michael and Schmidt, Burkhard",
      title         = "{LHCb Physics Motivations for the 2016 Heavy-ion LHC run}",
      institution   = "CERN",
      number  = "LHCb-PUB-2016-011",
      address       = "Geneva",
      year          = "2016",
      url           = "https://cds.cern.ch/record/2145943",
}
